\documentclass[reprint,aps,showpacs]{revtex4}

\usepackage{graphicx}

\begin{document}


\title{Finite size effects in the thermodynamics of a free neutral scalar field}

\author{A.S.~Parvan$^{1,2,3}$}

\affiliation{$^{1}$Bogoliubov Laboratory of Theoretical Physics, Joint Institute for Nuclear Research, Dubna, Russia}

\affiliation{$^{2}$Department of Theoretical Physics, Horia Hulubei National Institute of Physics and Nuclear Engineering, Bucharest-Magurele, Romania}

\affiliation{$^{3}$Institute of Applied Physics, Moldova Academy of Sciences, Chisinau, Republic of Moldova}

\begin{abstract}
The exact analytical lattice results for the partition function of the free neutral scalar field in one spatial dimension in both the configuration and the momentum space were obtained in the framework of the path integral method. The symmetric square matrices of the bilinear forms on the vector space of fields in both configuration space and momentum space were found explicitly. The exact lattice results for the partition function were generalized to the three-dimensional spatial momentum space and the main thermodynamic quantities were derived both on the lattice and in the continuum limit. The thermodynamic properties and the finite volume corrections to the thermodynamic quantities of the free real scalar field were studied. We found that on the finite lattice the exact lattice results for the free massive neutral scalar field agree with the continuum limit only in the region of small values of temperature and volume. However, at these temperatures and volumes the continuum physical quantities for both massive and massless scalar field deviate essentially from their thermodynamic limit values and recover them only at high temperatures or/and large volumes in the thermodynamic limit.
\end{abstract}

\pacs{11.10.Wx; 05.30.-d}

\maketitle

\section{Introduction}~\label{sec1}
The process of particle (hadron) production from the hot medium created in heavy ion collisions and hadron-hadron reactions at the LHC~\cite{Abelev13,Khachatryan10_1,Chatrchyan13,Kharlov13} and the RHIC~\cite{Arsene05,Adams06,Adare11} calls for the introduction of finite temperature and volume into studies of quantum field theory (QFT). However, most theoretical approaches used for the description of the thermodynamic properties or/and phase transitions for relativistic quantum fields (i.e. bosons and fermions, which are generated in these processes) are considered in the thermodynamic limit. In these models, the infinite volume of the system is implemented directly~\cite{Yagi,Cleymans86,BraunMunzinger09} or indirectly through the application of the continuous momentum spectra instead of the discrete eigenvalues of the momentum operator of particles~\cite{Stocker86,Andronic12,Kapoyannis07,Lee93,Kahara08,Shao11,Tiwari12}. Thus, the finite size effects are important for consideration in the statistical theories meant to describe the nucleus-nucleus and proton-proton collisions. They may significantly influence the thermodynamic characteristics of quantum fields and, in particular, of the QCD phase diagram as a whole~\cite{Braun13}.

There are various methods of quantization of fields, such as the canonical method, the path integral method, etc.~\cite{Greiner96,Kapusta89,Greiner07}. The thermodynamic properties of the QCD phase diagram are described by both the canonical quantization formalism (second quantization)~\cite{Shao11,Strodthoff12} and the path-integral formalism~\cite{Muta98}. Numerical computations in the framework of the path-integral formalism by the Monte Carlo method, for example in the case of lattice gauge theory, allow the study of nonperturbative effects, for example those manifesting themselves in the QCD phase transitions~\cite{Gattringer10}. All of these methods of field quantization are equivalent~\cite{Greiner96,Kapusta89}.

The formalism of statistical mechanics agrees with the zeroth law of thermodynamics under the condition that the thermodynamic potential of the statistical ensemble, which contains all information about the physical system, is a homogeneous function of the first order with respect to the extensive variables of state~\cite{Prigogine,Parvan1,Parvan2,Parvan3,Parvan2015,Parvan2015a}. For most physical systems this condition is fulfilled in the thermodynamic limit. The problem of homogeneity properties of the thermodynamic quantities in the case of lattice quantum chromodynamics (LQCD), which is expected to accurately describe the phase transitions for the relativistic quantum fields featuring in heavy ion collisions, has not been rigorously studied yet. It is considered that the trace anomaly enhancement is the interaction measure in both the pure gauge theory on the lattice~\cite{Boyd} and the full LQCD~\cite{Borsanyi14,Borsanyi10,Borsanyi13}. In the approximate numerical calculations for the lattice QCD as given, for example, in~\cite{Borsanyi10,Borsanyi13,Aoki06} it is difficult to verify the homogeneity properties of the thermodynamic potential of the system. Therefore, to investigate the severity of this problem, an exactly solvable statistical model for a free relativistic quantum field on a finite lattice should be considered and finite volume effects as well as the continuum limit should be carefully analyzed~\cite{Redlich}.

The main purpose of this paper is to obtain the analytical results for the partition function of the free neutral scalar field on a finite lattice in both configuration space and momentum space and to study the thermodynamic properties and the finite volume corrections to the thermodynamic quantities both on the finite lattice and in the continuum limit at finite temperature and volume.

The structure of the paper is as follows. In Section~\ref{sec2}, we briefly describe the path integral formalism for the neutral scalar field in the configuration and momentum spaces. The continuum limit for the thermodynamic quantities is given in Section~\ref{sec3}. The vacuum and physical thermodynamic quantities on the finite lattice are defined in Section~\ref{sec4}. The thermodynamic principle of additivity in the grand canonical ensemble is considered in Section~\ref{ap1}. The results are discussed in Section~\ref{sec5}. The main conclusions are summarized in the final section.

\section{Free real scalar field}~\label{sec2}
We consider a path integral method for the quantum fields to solve exactly the partition function for the free real scalar field in the one-spatial dimension, which is based on the discretization of the inverse temperature instead of time and uses the field operators in the Schr\"{o}dinger picture at time $t=0$. We also recite the most important path integral relations to keep the paper self-contained.

Let us consider a system consisting only of the neutral scalar field $\phi(x)$ with mass $m$. The corresponding classical action $S$ is the integral over space-time~\cite{Greiner96}
\begin{equation}\label{1}
  S=\int dt L = \int d^{4}x {\cal L}(\phi(x),\partial_{\mu}\phi(x)),
\end{equation}
with the classical Lagrangian density $\cal L$ given by
\begin{equation}\label{2}
  {\cal L} =\frac{1}{2} (\partial^{\mu}\phi \partial_{\mu}\phi - m^{2} \phi^{2}).
\end{equation}
Here and throughout the paper we use the system of natural units $\hbar=c=k_{B}=1$.

The thermodynamics of the real scalar field of volume $V$, in contact with the heat and particle reservoir of temperature $T$ and chemical potential $\mu$, is defined by the statistical operator and the partition function,
\begin{equation}\label{3}
  \hat{\varrho} =\frac{1}{Z} e^{-\beta (\hat{H}-\mu\hat{Q})} \quad \mathrm{and} \quad Z = Tr\ e^{-\beta (\hat{H}-\mu\hat{Q})},
\end{equation}
respectively, where $\beta=1/T$, $\hat{H}$ is the Hamiltonian of the system and $\hat{Q}$ is the electric charge operator in the canonical operator formalism. For the real scalar field the charge operator $\hat{Q}=0$.

Let us rewrite the partition function (\ref{3}) in the path integral method as~\cite{Kapusta89}
\begin{equation}\label{4}
    Z = \int [d\phi] \langle\phi | e^{-\beta (\hat{H}-\mu\hat{Q})} |\phi\rangle,
\end{equation}
where all the operators are defined in the Schr\"{o}dinger representation. The field operator $\hat{\phi}(x)$ in the Schr\"{o}dinger picture does not depend on time $t$, $\hat{\phi}(\vec{x})$ is $\hat{\phi}(x)$ at $t=0$ and has the eigenstate $|\phi(\vec{x})\rangle$ with the eigenvalue $\phi(\vec{x})$~\cite{Kapusta89}:
\begin{equation}\label{5}
  \hat{\phi}(\vec{x}) |\phi(\vec{x})\rangle = \phi(\vec{x}) |\phi(\vec{x})\rangle.
\end{equation}
Similar relations are satisfied by the canonical conjugate field operator $\hat{\pi}(x)=\partial \hat{{\cal L}}/\partial \dot{\hat{\phi}}(x) = \dot{\hat{\phi}}(x)$. The Hamiltonian $\hat{H}$ of the system in the Schr\"{o}dinger picture is given by~\cite{Kapusta89}
\begin{equation}\label{5a}
 \hat{H} = \int d^{3} x \hat{\mathcal{H}}(\hat{\pi}(\vec{x}),\hat{\phi}(\vec{x})),
\end{equation}
where the Hamiltonian density $\hat{\mathcal{H}}(\hat{\pi},\hat{\phi})$ is expressed in terms of the field operator $\hat{\phi}(\vec{x})$ and its conjugate momentum operator $\hat{\pi}(\vec{x})$:
\begin{equation}\label{5b}
  \hat{\mathcal{H}} = \frac{1}{2}\left[\hat{\pi}^{2} + (\vec{\nabla}\hat{\phi})^{2} + m^{2}\hat{\phi}^{2}\right].
\end{equation}

The main point of the path integral method considered here is that the ``time'' interval of continually variable length $\beta$ is subdivided into $N_{\beta}$ equal intervals of length $a_{\beta}$, $\beta=a_{\beta}N_{\beta}$, and the volume $V$ of the system is divided into $N_{\sigma}^{3}$ small cells, each with volume $a_{\sigma}^{3}$ so that $V=L_{\sigma}^{3}$, where $L_{\sigma}=a_{\sigma}N_{\sigma}$. The small intervals on the $\beta$-axis are labeled by the integer $n_{\beta}=1,\ldots,N_{\beta}$ and the space cells in $V$ are fixed by the integer vector $\vec{n}=(n_{x},n_{y},n_{z})$ with the coordinates $n_{\alpha}=1,\ldots,N_{\sigma}$. The discretized $\beta$ and volume $V$ form a $4$-dimensional lattice $\Lambda$, the cells of which are given by the vector $n_{\nu}=(\vec{n},n_{\beta})$. Then to each cell $n_{\nu}$ there is attributed the field operator $\hat{\phi}(n_{\nu})$ with an eigenstate $|\phi(n_{\nu})\rangle$ and eigenvalue $\phi(n_{\nu})$. We have a discrete set of a finite number of operators. For convenience, we denote the field operator in another form $\hat{\phi}_{l,i}\equiv\hat{\phi}(n_{\nu})$ ($l\in \Lambda_{3},i=n_{\beta}$), where $l$ is an integer and $\Lambda_{3}$ is the $3$-dimensional subspace of the lattice $\Lambda$.

For simplicity, to start with, let us find the analytical expression for the partition function in {\it one spatial dimension} and generalize it finally to the usual three spatial dimensions. The partition function (\ref{4}) in one spatial dimension can be written as
\begin{eqnarray}\label{6}
  Z = \lim_{N_{\sigma}\to\infty,N_{\beta}\to\infty} Z_{lat}
\end{eqnarray}
and
\begin{equation}\label{7}
  Z_{lat} = \int \prod_{l=1}^{N_{\sigma}} d\phi_{l,1} \ \langle \phi_{1,1},\ldots,\phi_{N_{\sigma},1} | e^{-a_{\beta} (\hat{H}-\mu\hat{Q})} e^{-a_{\beta} (\hat{H}-\mu\hat{Q})}\cdots e^{-a_{\beta} (\hat{H}-\mu\hat{Q})} | \phi_{1,1},\ldots,\phi_{N_{\sigma},1} \rangle,
\end{equation}
where $\lim$ denotes the continuum limit which means that $N_{\sigma}\to\infty,a_{\sigma}\to 0$ at $L_{\sigma}=const$ and $N_{\beta}\to \infty,a_{\beta}\to 0$ at $\beta=const$. The exponents in Eq.~(\ref{7}) are enumerated from right to left by the index $i=1,\ldots,N_{\beta}$. The state vectors $|\phi_{i}\rangle\equiv| \phi_{1,i},\ldots,\phi_{N_{\sigma},i} \rangle$ and $|\pi_{i}\rangle\equiv| \pi_{1,i},\ldots,\pi_{N_{\sigma},i} \rangle$ are orthogonal and complete:
\begin{eqnarray}\label{8}
    \langle \phi_{i} | \phi_{j}\rangle &=& \prod_{l=1}^{N_{\sigma}} \delta(\phi_{l,i}-\phi_{l,j}), \\ \label{8a}
    1 &=& \int \prod_{l=1}^{N_{\sigma}} d\phi_{l,i} \
    | \phi_{i} \rangle \langle \phi_{i} |
\end{eqnarray}
and
\begin{eqnarray}\label{9}
    \langle \pi_{i} | \pi_{j} \rangle &=&  \prod_{l=1}^{N_{\sigma}} \frac{2\pi}{a_{\sigma}} \delta(\pi_{l,i}-\pi_{l,j}), \\ \label{9a}
    1 &=& \int \prod_{l=1}^{N_{\sigma}} \frac{a_{\sigma}d\pi_{l,i}}{2\pi} \ | \pi_{i} \rangle \langle \pi_{i} |.
\end{eqnarray}
To solve Eq.~(\ref{7}), we insert in the left side of each $i$-th exponent the product of two unit operators for the fields $|\phi_{i+1}\rangle$ and $|\pi_{i}\rangle$. Using Eq.~(\ref{8}), we obtain
\begin{equation}\label{10}
Z_{lat} = \int\prod_{l=1}^{N_{\sigma}} d\phi_{l,1} \langle \phi_{1} | \phi_{N_{\beta}+1}\rangle \int\prod_{l=1}^{N_{\sigma}} \prod_{i=1}^{N_{\beta}} \frac{a_{\sigma}d\pi_{l,i}d\phi_{l,i}}{2\pi} \prod_{i=1}^{N_{\beta}} \langle \phi_{i+1} | \pi_{i} \rangle \langle \pi_{i} | e^{-a_{\beta} (\hat{H}-\mu\hat{Q})} | \phi_{i} \rangle.
\end{equation}

The matrix element is
\begin{equation}\label{10a}
   \langle \pi_{i} | e^{-a_{\beta} (\hat{H}-\mu\hat{Q})} | \phi_{i} \rangle = e^{-a_{\sigma}a_{\beta} \sum\limits_{l=1}^{N_{\sigma}} [\mathcal{H}_{l,i}-\mu\mathcal{Q}_{l,i}]}
   \langle \pi_{i} | \phi_{i} \rangle,
\end{equation}
where the Hamiltonian function (\ref{5b}) for the Klein-Gordon field on the lattice in one spatial dimension is given by
\begin{equation}\label{10b}
  \mathcal{H}_{l,i} = \frac{1}{2}\pi_{l,i}^{2} +\frac{1}{2} \left(\frac{\phi_{l+1,i}-\phi_{l,i}}{a_{\sigma}}\right)^{2} +\frac{1}{2} m^{2}\phi_{l,i}^{2}.
\end{equation}
Note that for the real scalar field the chemical potential $\mu=0$. The eigenstate $|\pi_{i} \rangle$ in the field representation can be written as~\cite{Greiner96}
\begin{equation}\label{10bc}
  \langle \pi_{i} | \phi_{i} \rangle=e^{-\imath a_{\sigma}\sum\limits_{l=1}^{N_{\sigma}} \pi_{l,i} \phi_{l,i}}.
\end{equation}

Substituting Eqs.~(\ref{8}), (\ref{10a}), (\ref{10b}) and (\ref{10bc}) into Eq.~(\ref{10}) and introducing the (anti)periodic boundary conditions for the field $\phi_{l,i}$ in the one spatial dimension, we obtain
\begin{equation}\label{11}
   Z_{lat} = \int \prod_{l=1}^{N_{\sigma}} \prod_{i=1}^{N_{\beta}} \frac{a_{\sigma}d\pi_{l,i}d\phi_{l,i}}{2\pi} \ e^{-a_{\sigma}a_{\beta}\sum\limits_{l=1}^{N_{\sigma}}\sum\limits_{i=1}^{N_{\beta}} \left[\frac{1}{2}\pi_{l,i}^{2}  - \imath \pi_{l,i}\left(\frac{\phi_{l,i+1}-\phi_{l,i}}{a_{\beta}}\right) +\frac{1}{2} \left(\frac{\phi_{l+1,i}-\phi_{l,i}}{a_{\sigma}}\right)^{2} +\frac{1}{2} m^{2}\phi_{l,i}^{2} \right]}
\end{equation}
with the conditions that
\begin{equation}\label{11a}
 \phi_{l,N_{\beta}+1} = \phi_{l,1} \quad \mathrm{and} \quad \phi_{N_{\sigma}+1,i} = \xi_{\sigma} \phi_{1,i},
\end{equation}
where $\xi_{\sigma}=\pm 1$ for the periodic and antiperiodic boundary conditions along the spatial $x$-axis.

Integrating Eq.~(\ref{11}) over the variable $\pi_{l,i}$ under the condition that $\mathrm{Re}(a_{\sigma}a_{\beta}/2)>0$, we obtain
\begin{equation}\label{11b}
   Z_{lat} = \left(\frac{a_{\sigma}}{2\pi a_{\beta}}\right)^{\frac{N_{\sigma}N_{\beta}}{2}} \int \prod_{l=1}^{N_{\sigma}} \prod_{i=1}^{N_{\beta}} d\phi_{l,i} \ e^{-a_{\sigma}a_{\beta}\sum\limits_{l=1}^{N_{\sigma}}\sum\limits_{i=1}^{N_{\beta}} \left[\frac{1}{2}\left(\frac{\phi_{l,i+1}-\phi_{l,i}}{a_{\beta}}\right)^{2} +\frac{1}{2} \left(\frac{\phi_{l+1,i}-\phi_{l,i}}{a_{\sigma}}\right)^{2} +\frac{1}{2} m^{2}\phi_{l,i}^{2} \right]}
\end{equation}
with the restrictions given in Eq.~(\ref{11a}). Equation (\ref{11b}) exactly coincides with Eq.~(2.7) from Ref.~\cite{Engels} in the case of the one spatial dimension. Thus, the path integral method used in this paper gives the same results for the partition function of the scalar field as the standard method of path integrals.

We write the function in the exponent in Eq.~(\ref{11b}) into the bilinear form on the $n$ dimensional vector space of vectors $\vec{\Phi}=(\Phi_{1},\ldots,\Phi_{n})$ under the conditions (\ref{11a}), where $n=N_{\sigma}N_{\beta}$ and  $\Phi_{J(l,i)}=(a_{\sigma}/2a_{\beta})^{1/2}\phi_{l,i}$ with $J(l,i)=N_{\beta}(l-1)+i$. We have
\begin{equation}\label{12}
Z_{lat} = \pi^{-\frac{N_{\sigma}N_{\beta}}{2}} \int\limits_{-\infty}^{\infty} \prod_{l=1}^{N_{\sigma}} \prod_{i=1}^{N_{\beta}}d\Phi_{J(l,i)} \
e^{-\sum\limits_{l=1}^{N_{\sigma}}\sum\limits_{i=1}^{N_{\beta}} \sum\limits_{k=1}^{N_{\sigma}}\sum\limits_{j=1}^{N_{\beta}} A_{J(l,i) J(k,j)} \Phi_{J(l,i)} \Phi_{J(k,j)}}
\end{equation}
with
\begin{eqnarray}\label{13}\nonumber
 A_{J(l,i) J(k,j)} &=&  \delta_{l,k}\delta_{i,j} \ \left[2 \left(1+\frac{a_{\beta}^{2}}{a_{\sigma}^{2}}\right)+a_{\beta}^{2}m^{2}\right] -  \delta_{l,k} \ [(1-\delta_{i,N_{\beta}})\ \delta_{j,i+1}+(1-\delta_{j,N_{\beta}})\ \delta_{i,j+1}+\delta_{i,1}\delta_{j,N_{\beta}}+\delta_{i,N_{\beta}}\delta_{j,1}]
  \nonumber \\    &-&  \delta_{i,j}  \left[(1-\delta_{l,N_{\sigma}})\ \delta_{k,l+1}+(1-\delta_{k,N_{\sigma}})\ \delta_{l,k+1}+ \xi_{\sigma}   (\delta_{l,1}\delta_{k,N_{\sigma}}+\delta_{l,N_{\sigma}}\delta_{k,1}) \right] \frac{a_{\beta}^{2}}{a_{\sigma}^{2}},
\end{eqnarray}
where $J=1,2,\ldots,n$ is the number of the site on the lattice in the configuration space. The matrix $A$ is a symmetric square matrix of size $n\times n$, which for the periodic boundary conditions along the spatial $x$-axis can be written as
\begin{equation}\label{12a}
  A= I_{N_{\sigma}} \otimes I_{N_{\beta}} \left[2 \left(1+\frac{a_{\beta}^{2}}{a_{\sigma}^{2}}\right)+a_{\beta}^{2}m^{2}\right]  - I_{N_{\sigma}} \otimes C_{N_{\beta}}  -
  \frac{a_{\beta}^{2}}{a_{\sigma}^{2}} C_{N_{\sigma}} \otimes I_{N_{\beta}} ,
\end{equation}
where $I_{N}$ is the identity matrix of size $N\times N$, the symbol $\otimes$ denotes the tensor product and $C_{N}$ is the matrix of size $N\times N$ of the form
\begin{equation}\label{12b}
  C_{N} = \left(
            \begin{array}{cccccccc}
              0 & 1 & 0 & 0 & \cdots & 0 & 0 & 1 \\
              1 & 0 & 1 & 0 & \cdots & 0 & 0 & 0 \\
              0 & 1 & 0 & 1 & \cdots & 0 & 0 & 0 \\
              0 & 0 & 1 & 0 & \cdots & 0 & 0 & 0 \\
              \vdots & \vdots & \vdots & \vdots & \ddots & \vdots & \vdots & \vdots \\
              0 & 0 & 0 & 0 & \cdots & 0 & 1 & 0 \\
              0 & 0 & 0 & 0 & \cdots & 1 & 0 & 1 \\
              1 & 0 & 0 & 0 & \cdots & 0 & 1 & 0 \\
            \end{array}
          \right).
\end{equation}

For any real symmetric square matrix $D$ the following formula for Riemann integrals is valid~\cite{Kapusta89}:
\begin{equation}\label{13a}
  \int\limits_{-\infty}^{\infty} dx_{1}\cdots dx_{n} e^{-x_{i}D_{ij}x_{j}}=\pi^{n/2} (\det D)^{-1/2}.
\end{equation}
Thus, using (\ref{13a}) and integrating (\ref{12}) with respect to the variables $\Phi_{J(l,i)}$, we obtain
\begin{equation}\label{14}
  Z_{lat} = \frac{1}{\sqrt{\det A}} = \prod_{l=1}^{N_{\sigma}N_{\beta}} P_{l}^{-1/2}
\end{equation}
with the recurrence equations of the form
\begin{eqnarray}\label{16}
  P_{l}   &=& A_{ll}-\frac{1}{4}\sum\limits_{k=1}^{l-1} \frac{Q_{kl}^{2}}{P_{k}},    \\ \label{17}
  Q_{ij}  &=& 2 A_{ij}- \frac{1}{2}\sum\limits_{k=1}^{i-1} \frac{Q_{ki}Q_{kj}}{P_{k}},
\end{eqnarray}
where $\mathrm{Re} (P_{l})>0$, $l=1,\ldots,N_{\sigma}N_{\beta}$, $i=1,\ldots,N_{\sigma}N_{\beta}-1$ and $j=i+1,\ldots,N_{\sigma}N_{\beta}$. The matrix elements $A_{ij}$ of the matrix $A$ are given in Eq.~(\ref{13}). Note that Eqs.~(\ref{16}) and (\ref{17}) can be reduced to the equations for the LDL factorization~\cite{Watkins} by changing the variables.

Let us rewrite the partition function (\ref{7}) in the momentum space on the basis of the Fourier transform defined on the lattice. The four-dimensional momentum space $\tilde{\Lambda}$, which corresponds to the lattice $\Lambda$, can be defined as
\begin{eqnarray}\label{18}
    \tilde{\Lambda}&=&\left\{p_{\mu}=(p_{x},p_{y},p_{z},p_{\beta})\left| p_{\mu} =\frac{2\pi}{a_{\mu}N_{\mu}} (k_{\mu}+\theta_{\mu}), k_{\mu 1}\leq k_{\mu}\leq k_{\mu 2}, \mu=x,y,z,\beta \right.\right\}
\end{eqnarray}
and
\begin{eqnarray}\label{19}
     k_{\mu 1}&=&-\frac{N_{\mu}-1}{2}+ \frac{\eta_{\mu}\xi_{\mu}}{2}, \\ \label{19a}
     k_{\mu 2}&=&\frac{N_{\mu}-1}{2}+ \frac{\eta_{\mu}\xi_{\mu}}{2},
\end{eqnarray}
where $\xi_{\mu}=1$ $(\theta_{\mu}=0)$ and $\xi_{\mu}=-1$ $(\theta_{\mu}=1/2)$ for the choice of periodic and antiperiodic boundary conditions, respectively, along the $\mu$-direction and $\eta_{\mu}=1$ for $N_{\mu}$ even and $\eta_{\mu}=0$ for $N_{\mu}$ odd~\cite{Gattringer10}.

It is not difficult to prove that the vectors $p_{\mu}$ and $n_{\mu}$ satisfy the following relations~\cite{Gattringer10}:
\begin{eqnarray}\label{20}
 \frac{1}{N_{\mu}} \sum\limits_{n_{\mu}=1}^{N_{\mu}} e^{\imath \frac{2\pi}{N_{\mu}} (k_{\mu}-k'_{\mu})n_{\mu}} &=& \delta_{k_{\mu},k'_{\mu}}  , \\ \label{21}
 \frac{1}{N_{\mu}} \sum\limits_{k_{\mu}=k_{\mu 1}}^{k_{\mu 2}} e^{\imath \frac{2\pi}{N_{\mu}} (n_{\mu}-n'_{\mu})(k_{\mu}+\theta_{\mu})} &=& \delta_{n_{\mu},n'_{\mu}}
\end{eqnarray}
and
\begin{equation}\label{22}
   \frac{1}{N_{\mu}} \sum\limits_{n_{\mu}=1}^{N_{\mu}} e^{\imath a_{\mu} (p_{\mu}+p'_{\mu})n_{\mu}} = \delta_{k_{\mu}+k'_{\mu}+2\theta_{\mu},0} + [(1-2\theta_{\mu})\eta_{\mu}+2\theta_{\mu}(1-\eta_{\mu})] \delta_{k_{\mu}+k'_{\mu}+2\theta_{\mu},N_{\mu}},
\end{equation}
where $-(N_{\mu}-1)\leq k_{\mu}-k'_{\mu}\leq N_{\mu}-1$ and $-(N_{\mu}-1)\leq n_{\mu}-n'_{\mu}\leq N_{\mu}-1$.

Then the Fourier transform for the lattice field $\Phi_{J(n_{x},n_{\beta})}$ and its inverse transform can be written as~\cite{Gattringer10}
\begin{eqnarray}\label{23}
  \Phi_{J(n_{x},n_{\beta})} &=& \frac{1}{\sqrt{N_{\sigma}N_{\beta}}}  \sum\limits_{k_{x}=k_{x 1}}^{k_{x 2}} \sum\limits_{k_{\beta}=k_{\beta 1}}^{k_{\beta 2}} f_{I(k_{x},k_{\beta})} e^{\imath (a_{\sigma}p_{x} n_{x}+a_{\beta}p_{\beta}n_{\beta})},  \\ \label{24}
  f_{I(k_{x},k_{\beta})} &=& \frac{1}{\sqrt{N_{\sigma}N_{\beta}}} \sum\limits_{n_{x}=1}^{N_{\sigma}} \sum\limits_{n_{\beta}=1}^{N_{\beta}} \Phi_{J(n_{x},n_{\beta})}  e^{-\imath (a_{\sigma}p_{x}n_{x}+a_{\beta}p_{\beta}n_{\beta})},
\end{eqnarray}
where the index $I(k_{x},k_{\beta})\equiv N_{\beta} (k_{x}-k_{x 1})+k_{\beta}-k_{\beta 1}+1$ is the number of the site on the lattice in the momentum space. It should be stressed that the values of the indices $J$ and $I$ are fixed by the vectors $n_{\mu}$ and $k_{\mu}$, respectively. The neutral scalar field is real and $\Phi_{J}^{*}=\Phi_{J}$. Therefore, the complex function $f_{I(k_{x},k_{\beta})}$ can be represented by its amplitude and phase in the form
\begin{equation}\label{25}
    f_{I(k_{x},k_{\beta})} = R_{I(k_{x},k_{\beta})} \ e^{\imath (a_{\sigma}p_{x}+a_{\beta}p_{\beta})}.
\end{equation}
Substituting Eqs.~(\ref{25}) and (\ref{23}) into (\ref{12}) and using Eq.~(\ref{22}), we can write
\begin{eqnarray}\label{26}
   Z_{lat} &=& \pi^{-\frac{N_{\sigma}N_{\beta}}{2}} |\det \mathcal{J}| \int\limits_{-\infty}^{\infty} \prod_{k_{x}=k_{x 1}}^{k_{x 2}} \prod_{k_{\beta}=k_{\beta 1}}^{k_{\beta 2}} dR_{I(k_{x},k_{\beta})} \nonumber \\
   &\times& e^{-\sum\limits_{k_{x}=k_{x 1}}^{k_{x 2}}\sum\limits_{k_{\beta}=k_{\beta 1}}^{k_{\beta 2}} \sum\limits_{k'_{x}=k_{x 1}}^{k_{x 2}}\sum\limits_{k'_{\beta}=k_{\beta 1}}^{k_{\beta 2}} \  B_{I(k_{x},k_{\beta})I(k'_{x},k'_{\beta})} R_{I(k_{x},k_{\beta})} R_{I(k'_{x},k'_{\beta})}}
\end{eqnarray}
and
\begin{equation}\label{28}
  B_{I(k_{x},k_{\beta})I(k'_{x},k'_{\beta})} =  G_{I(k_{x},k_{\beta})}\ (\delta_{k_{\beta}+k'_{\beta},0} +\eta_{\beta} \delta_{k_{\beta}+k'_{\beta},N_{\beta}}) \left(\delta_{k_{x}+k'_{x}+2\theta_{x},0} + [(1-2\theta_{x})\eta_{x}+2\theta_{x}(1-\eta_{x})] \delta_{k_{x}+k'_{x}+2\theta_{x},N_{\sigma}} \right), \;\;\;\;
\end{equation}
where
\begin{equation}\label{30}
  G_{I(k_{x},k_{\beta})} =  a_{\beta}^{2}\left[\left(\frac{2}{a_{\beta}} \sin\frac{a_{\beta}p_{\beta}}{2}\right)^{2}
+ \left(\frac{2}{a_{\sigma}} \sin\frac{a_{\sigma}p_{x}}{2}\right)^{2} + m^{2} \right]
\end{equation}
and
\begin{equation}\label{29}
\mathcal{J}_{J(n_{x},n_{\beta})I(k_{x},k_{\beta})} = \frac{\partial \Phi_{J(n_{x},n_{\beta})}}{\partial R_{I(k_{x},k_{\beta})}} =
\frac{1}{\sqrt{N_{\sigma}N_{\beta}}} \  e^{\imath [a_{\sigma}p_{x} (n_{x}+1)+a_{\beta}p_{\beta}(n_{\beta}+1)]}.
\end{equation}
Here $\mathcal{J}_{J(n_{x},n_{\beta})I(k_{x},k_{\beta})}$ are the matrix elements of the Jacobian matrix $\mathcal{J}$ of size $n\times n$. The matrix $B$ is a symmetric square matrix of size $n\times n$, which for the periodic boundary conditions along the spatial $x$-axis in the case of $N_{\sigma}$ odd and $N_{\sigma}$ even can be written in a block form as
\begin{equation}\label{29a}
   B = \left(
            \begin{array}{cccccc}
              0 & 0 &  \cdots & 0 & 0 & B_{1} \\
              0 & 0 &  \cdots & 0 & B_{2} & 0 \\
              0 & 0 &  \cdots & B_{3} & 0 & 0 \\
             \vdots & \vdots & \ddots & \vdots & \vdots & \vdots \\
             0 & 0 &  \cdots & 0 & 0 & 0 \\
              0 & B_{N_{\sigma}-1} &  \cdots & 0 & 0 & 0 \\
              B_{N_{\sigma}} & 0 &  \cdots & 0 & 0 & 0 \\
            \end{array}
          \right)
          \quad \mathrm{and} \quad
   B = \left(
            \begin{array}{cccccc}
              0 & 0 &  \cdots & 0 & B_{1} & 0 \\
              0 & 0 &  \cdots & B_{2} & 0 & 0 \\
              0 & 0 &  \cdots & 0 & 0 & 0 \\
             \vdots & \vdots & \ddots & \vdots & \vdots & \vdots \\
              0 & B_{N_{\sigma}-2} &  \cdots & 0 & 0 & 0 \\
              B_{N_{\sigma}-1} & 0 &  \cdots & 0 & 0 & 0 \\
              0 & 0 &  \cdots & 0 & 0 &  B_{N_{\sigma}} \\
            \end{array}
          \right),
\end{equation}
respectively, where the $i$-th matrix $B_{i}$ is a symmetric square matrix of size $N_{\beta}\times N_{\beta}$, which in the case of $N_{\beta}$ odd and $N_{\beta}$ even are given by
\begin{equation}\label{29b}
   B_{i} = \left(
            \begin{array}{cccccc}
              0 & 0 &  \cdots & 0 & 0 & G_{j+1} \\
              0 & 0 &  \cdots & 0 & G_{j+2} & 0 \\
              0 & 0 &  \cdots &G_{j+3} & 0 & 0 \\
             \vdots & \vdots & \ddots & \vdots & \vdots & \vdots \\
              0 & 0 &  \cdots & 0 & 0 & 0 \\
              0 & G_{j+N_{\beta}-1} &  \cdots & 0 & 0 & 0 \\
              G_{j+N_{\beta}} & 0 &  \cdots & 0 & 0 & 0 \\
            \end{array}
          \right)
\end{equation}
and
\begin{equation}
    B_{i} = \left(
            \begin{array}{cccccc}
              0 & 0 &  \cdots & 0 & G_{j+1} & 0 \\
              0 & 0 &  \cdots & G_{j+2} & 0 & 0 \\
              0 & 0 &  \cdots & 0 & 0 & 0 \\
             \vdots & \vdots & \ddots & \vdots & \vdots & \vdots \\
              0 & G_{j+N_{\beta}-2} &  \cdots & 0 & 0 & 0 \\
              G_{j+N_{\beta}-1} & 0 &  \cdots & 0 & 0 & 0 \\
              0 & 0 &  \cdots & 0 & 0 &  G_{j+N_{\beta}} \\
            \end{array}
          \right),
\end{equation}
respectively. Here we have $j=N_{\beta}(i-1)$. Note that the arguments of the composed index $I(k_{x},k_{\beta})$ of $G_{I(k_{x},k_{\beta})}$ can be found from its definition given above.

Using (\ref{13a}) and integrating (\ref{26}) with respect to the variables $R_{I(k_{x},k_{\beta})}$, we obtain
\begin{equation}\label{31}
    Z_{lat} = \frac{|\det \mathcal{J}|}{\sqrt{\det B}}.
\end{equation}
The matrix $B$ has $N_{\sigma}N_{\beta}$ nonzero elements equal to $G_{I(k_{x},k_{\beta})}$. In any row and any column there is only one nonzero element, regardless of the periodicity conditions for the $x$-axis and parity conditions for $N_{\sigma}$ and $N_{\beta}$. Therefore, the determinant of the matrix $B$ can be written as
\begin{eqnarray}\label{32}
    \det B &=& \chi \ \prod_{k_{x}=k_{x 1}}^{k_{x 2}} \prod_{k_{\beta}=k_{\beta 1}}^{k_{\beta 2}} \  G_{I(k_{x},k_{\beta})}, \\ \label{33}
     \chi &=& (-1)^{\frac{1}{2}N_{\sigma}I_{\beta}+\frac{1}{2}N_{\beta}^{2} [(N_{\sigma}-I_{x})^{2}+I_{x}]},
\end{eqnarray}
where $I_{\beta} = (1-\eta_{\beta}) 3 N_{\beta} +\eta_{\beta} (N_{\beta}+2)$ and $I_{x} = (1-\eta_{x}) 2 \theta_{x} +\eta_{x} (1-2\theta_{x})$. Substituting Eq.~(\ref{32}) into Eq.~(\ref{31}), we obtain
\begin{equation}\label{34}
    Z_{lat} =  \frac{1}{\sqrt{\prod_{k_{x}=k_{x 1}}^{k_{x 2}} \prod_{k_{\beta}=k_{\beta 1}}^{k_{\beta 2}} \  G_{I(k_{x},k_{\beta})}}},
\end{equation}
where $|\det \mathcal{J}|\chi^{-1/2}=1$ and $G_{I(k_{x},k_{\beta})}$ is the function calculated by formula (\ref{30}). Note that Eq.~(\ref{34}) is equivalent to Eq.~(\ref{14}). For example, calculating $Z_{lat}$ by Eqs.~(\ref{34}) and (\ref{14}) for $N_{\sigma}=N_{\beta}=4$, $L_{\sigma}=9$ fm at temperature $T=100$ MeV and $m=134.98$ MeV, we obtain exactly the same numerical result, $Z_{lat}=0.0604$.

Now the partition function (\ref{34}) for one spatial dimension can be generalized to {\it three spatial dimensions} of the momentum space. For this reason, we can rewrite the partition function (\ref{34}) and the function (\ref{30}) in the form
\begin{eqnarray}\label{35}
    Z_{lat} &=& \prod_{\vec{k}=k_{\sigma 1}}^{k_{\sigma 2}} \prod_{k_{\beta}=k_{\beta 1}}^{k_{\beta 2}} \ \frac{1}{\sqrt{a_{\beta}^{2}(\omega_{\beta}^{2} + \omega^{2})}},  \\ \label{35a}
    \omega_{\beta} &=& \frac{2}{a_{\beta}} \sin\frac{a_{\beta}p_{\beta}}{2},    \\ \label{36}
  \omega &=& \sqrt{\sum\limits_{\alpha=1}^{3} \left(\frac{2}{a_{\sigma}} \sin\frac{a_{\sigma}p_{\alpha}}{2}\right)^{2} + m^{2}},
\end{eqnarray}
where $\omega$ is a relativistic one-particle energy on the lattice, $\theta_{\beta}=0$ and $p_{\beta}=2\pi k_{\beta}/\beta$ is the Matsubara frequency. Let us remark that the lattice partition function (\ref{35}) for the particular case of the periodic spatial boundary conditions, $\theta_{\sigma}=0$, was firstly obtained by another method in~\cite{Engels}.

The thermodynamic quantities on the lattice are derived from the partition function (\ref{35}). The density of the thermodynamic potential $\omega_{E}=-(\beta V)^{-1}\ln Z_{lat}$, the energy density $\varepsilon_{E}=-V^{-1}\partial\ln Z_{lat}/\partial\beta$, the pressure $p_{E}=\beta^{-1}\partial \ln Z_{lat}/\partial V$, and the entropy density $s_{E}$ can be written as
\begin{eqnarray}\label{37}
  \omega_{E} &=& \frac{1}{2\beta V}\sum\limits_{\vec{k}=k_{\sigma 1}}^{k_{\sigma 2}} \sum\limits_{k_{\beta}=k_{\beta 1}}^{k_{\beta 2}} \ln\left[a_{\beta}^{2}(\omega_{\beta}^{2} + \omega^{2})\right],   \\ \label{38}
  \varepsilon_{E} &=& \frac{1}{\beta V}\sum\limits_{\vec{k}=k_{\sigma 1}}^{k_{\sigma 2}} \sum\limits_{k_{\beta}=k_{\beta 1}}^{k_{\beta 2}} \frac{\omega^{2}}{\omega_{\beta}^{2} + \omega^{2}}, \\ \label{39}
  p_{E} &=&  \frac{1}{3\beta V}\sum\limits_{\vec{k}=k_{\sigma 1}}^{k_{\sigma 2}} \sum\limits_{k_{\beta}=k_{\beta 1}}^{k_{\beta 2}} \frac{\omega^{2}-m^{2}}{\omega_{\beta}^{2} + \omega^{2}}
\end{eqnarray}
and
\begin{equation}\label{40}
  s_{E} = \beta (-\omega_{E}+\varepsilon_{E}),
\end{equation}
where $\partial N_{\beta}/\partial\beta=0$ and $\partial N_{\sigma}/\partial V=0$. Note that the energy density (\ref{38}) and pressure (\ref{39}) with the periodic spatial boundary conditions, $\theta_{\sigma}=0$, were obtained in~\cite{Engels}.

For $m=0$, the zero-mode term $\vec{k}=0,k_{\beta}=0$ does not contribute to the sums in Eqs.~(\ref{37})-(\ref{39})~\cite{Engels}. Thus, for the free massless real scalar field we have the relation $p_{E}=\varepsilon_{E}/3$.

\section{Continuum limit}\label{sec3}
Let us find the density of the thermodynamic potential (\ref{37}), the energy density (\ref{38}) and the pressure (\ref{39}) in the continuum limit. Here we consider only the periodic boundary conditions, $\theta_{\sigma}=0$ and $\theta_{\beta}=0$, as in Ref.~\cite{Engels}.

First, let us find the thermodynamic quantities (\ref{37})--(\ref{39}) in the limit $N_{\beta}\to\infty,a_{\beta}\to 0$ at $\beta=const$ and fixed value of $N_{\sigma}$. Here and throughout the paper such thermodynamic quantities will be denoted by the asterisk symbol. Expanding the function under the sums in Eq.~(\ref{38}) into the series on $N_{\beta}^{-1}$ around the point $N_{\beta}=\infty$ and taking the limit $N_{\beta}\to\infty$, we obtain
\begin{equation}\label{42}
    \varepsilon_{E}^{*} =  \frac{1}{\beta V} \ \sum\limits_{\vec{k}=k_{\sigma 1}}^{k_{\sigma 2}} \sum\limits_{k_{\beta}=-\infty}^{\infty} \frac{(\beta\omega)^{2}}{\left(2\pi k_{\beta}\right)^{2} +(\beta\omega)^{2}},
\end{equation}
where $k_{\sigma 1}$ and $k_{\sigma 2}$ are defined in Eqs.~(\ref{19}) and (\ref{19a}), respectively, and $\omega$ is given in Eq.~(\ref{36}). Using the generating function (see, for example,~\cite{Gradshteyn})
\begin{eqnarray}\label{43}
  \sum\limits_{k=-\infty}^{\infty}  \frac{a^{2}}{(2 \pi k)^{2}+a^{2}} &=& \frac{a}{2} \coth\frac{a}{2},
\end{eqnarray}
one obtains
\begin{eqnarray}\label{44}
   \varepsilon_{E}^{*} &=& \frac{1}{\beta V} \sum\limits_{\vec{k}=k_{\sigma 1}}^{k_{\sigma 2}} \frac{\beta\omega}{2} \coth\frac{\beta\omega}{2} =  \varepsilon_{v}^{*}+ \varepsilon^{*}, \\ \label{45}
  \varepsilon_{v}^{*} &=& \frac{1}{2V} \sum\limits_{\vec{k}=k_{\sigma 1}}^{k_{\sigma 2}} \omega, \\ \label{46}
  \varepsilon^{*} &=& \frac{1}{V} \sum\limits_{\vec{k}=k_{\sigma 1}}^{k_{\sigma 2}} \omega \ \frac{1}{e^{\beta \omega}-1},
\end{eqnarray}
where $\varepsilon_{v}^{*}$ and $ \varepsilon^{*}$ are the vacuum and physical terms, respectively, of the energy density $\varepsilon_{E}^{*}$.

Using Eq.~(\ref{43}), the pressure (\ref{39}) in this limit can be rewritten as
\begin{eqnarray}\label{49}
   p_{E}^{*} &=& p_{v}^{*}+ p^{*}, \\ \label{50}
   p_{v}^{*} &=& \frac{1}{6V} \sum\limits_{\vec{k}=k_{\sigma 1}}^{k_{\sigma 2}} \left(\omega-\frac{m^{2}}{\omega}\right), \\ \label{51}
   p^{*} &=& \frac{1}{3V} \sum\limits_{\vec{k}=k_{\sigma 1}}^{k_{\sigma 2}} \left(\omega-\frac{m^{2}}{\omega}\right) \ \frac{1}{e^{\beta \omega}-1},
\end{eqnarray}
where $p_{v}^{*}$ and $p^{*}$ are the vacuum and physical terms, respectively, of the pressure $p_{E}^{*}$.

Considering Eqs.~(\ref{44}), (\ref{49}) and the definitions of the quantities (\ref{37})--(\ref{39}), the density of thermodynamic potential (\ref{37}) in this limit can be rewritten as
\begin{eqnarray}\label{55}
   \omega_{E}^{*}  &=& \omega_{v}^{*} + \omega_{ph}^{*}, \\ \label{56}
   \omega_{v}^{*} &=& \frac{1}{2V} \sum\limits_{\vec{k}=k_{\sigma 1}}^{k_{\sigma 2}} \omega, \\ \label{57}
   \omega_{ph}^{*} &=& \frac{1}{\beta V} \sum\limits_{\vec{k}=k_{\sigma 1}}^{k_{\sigma 2}} \ln\left(1-e^{-\beta\omega}\right),
\end{eqnarray}
where $\omega_{v}^{*}$ and $\omega_{ph}^{*}$ are the vacuum and physical terms, respectively, of the density of the thermodynamic potential $\omega_{E}^{*}$. Note that in the limit $N_{\beta}\to\infty,\beta=const$ and $N_{\sigma}=const$ all the vacuum and physical terms are not divergent. The numerical proof of Eq.~(\ref{55}) will be given in Section~\ref{sec5}.

Now let us find the thermodynamic quantities (\ref{37})--(\ref{39}) in the continuum limit. The thermodynamic quantities in the continuum limit will be denoted by the index 'c'. The energy density (\ref{44}) in the continuum limit can be rewritten as~\cite{Kapusta89,Bellac}
\begin{eqnarray}\label{58}
  \varepsilon_{E}^{c} &=&  \varepsilon_{v}^{c}+ \varepsilon^{c}, \\ \label{59}
  \varepsilon_{v}^{c} &=& \frac{1}{2V} \sum\limits_{\vec{k}} \omega, \\ \label{60}
  \varepsilon^{c} &=& \frac{1}{V} \sum\limits_{\vec{k}} \omega \ \frac{1}{e^{\beta \omega}-1},
\end{eqnarray}
where $k_{\alpha} = 0,\pm 1, \pm 2,\ldots\pm \infty$ $(\alpha=1,2,3)$ and the one-particle energy (\ref{36}) takes the form
\begin{equation}\label{61}
  \omega = \sqrt{\vec{p}^{2}+m^{2}}, \qquad  p_{\alpha}=\frac{2\pi}{L_{\sigma}}\ k_{\alpha}.
\end{equation}

The pressure (\ref{49}) in the continuum limit can be rewritten as~\cite{Kapusta89,Bellac}
\begin{eqnarray}\label{62}
   p_{E}^{c} &=& p_{v}^{c}+ p^{c}, \\ \label{63}
   p_{v}^{c} &=& \frac{1}{6V} \sum\limits_{\vec{k}} \left(\omega-\frac{m^{2}}{\omega}\right), \\ \label{64}
   p^{c} &=& \frac{1}{3V} \sum\limits_{\vec{k}} \left(\omega-\frac{m^{2}}{\omega}\right) \ \frac{1}{e^{\beta \omega}-1},
\end{eqnarray}
where $\omega$ is given in Eq.~(\ref{61}) and the range of $k_{\alpha}$ is given below Eq.~(\ref{60}).

The density of the thermodynamic potential (\ref{55}) in the continuum limit can be rewritten as~\cite{Kapusta89,Bellac}
\begin{eqnarray}\label{65}
   \omega_{E}^{c} &=& \omega_{v}^{c} + \omega_{ph}^{c}, \\ \label{66}
   \omega_{v}^{c} &=& \frac{1}{2V} \sum\limits_{\vec{k}} \omega =\varepsilon_{v}^{c}, \\ \label{67}
   \omega_{ph}^{c} &=& \frac{1}{\beta V} \sum\limits_{\vec{k}} \ln\left(1-e^{-\beta\omega}\right),
\end{eqnarray}
where $\omega$ is defined in Eq.~(\ref{61}) and the range of $k_{\alpha}$ is given below Eq.~(\ref{60}).  Thus, in the continuum limit the energy density (\ref{58})--(\ref{60}), the pressure (\ref{62})--(\ref{64}), and the density of the thermodynamic potential (\ref{65})--(\ref{67}) exactly coincide with their corresponding quantities obtained by the method of second quantization~\cite{Kapusta89,Bellac}. Hence, it was analytically proved that for the free neutral scalar field the method of path integral quantization and the method of canonical quantization are equivalent at any values of $T$ and $V$. Note that in the continuum limit all the vacuum terms are divergent.

Using Eqs.~(\ref{60}) and (\ref{64}) we can write the trace anomaly in the continuum limit as
\begin{equation}\label{tt1}
  \Delta_{2}^{c}\equiv \beta^{4} (\varepsilon^{c}-3p^{c}) = \frac{\beta^{4}}{V} \sum\limits_{\vec{k}}\frac{m^{2}}{\omega}\frac{1}{e^{\beta \omega}-1}.
\end{equation}
The trace anomaly $\Delta_{2}^{c}$ defines the deviation of the equation of state from the Stefan-Boltzmann equation. The trace anomaly (\ref{tt1}) for the massive neutral scalar field in a finite volume and at low temperatures is nonvanishing; however, it decreases with $T$.

The potential inhomogeneity in the continuum limit can be defined as
\begin{equation}\label{pp1}
  \Delta_{1}^{c} \equiv 3\beta^{4} (\omega_{ph}^{c}+p^{c}) = \frac{3\beta^{3}}{V} \sum\limits_{\vec{k}} \ln\left(1-e^{-\beta\omega}\right) +
    \frac{\beta^{4}}{V} \sum\limits_{\vec{k}}(\omega-\frac{m^{2}}{\omega})\frac{1}{e^{\beta \omega}-1}.
\end{equation}
This quantity is a measure of violation of the homogeneous properties of the thermodynamic potential of the grand canonical ensemble. The thermodynamic potential of the grand canonical ensemble is not a homogeneous function of the first order if the quantity (\ref{pp1}) is not equal to zero. For the proof of this statement see Section~\ref{ap1}.

For $m=0$, the zero-mode term $\vec{k}=0$ in all the equations given above is suppressed because this term describes a condensate of massless bosons which is unobservable. Thus, for the free massless real scalar field we have the relations $p_{E}^{*}=\varepsilon_{E}^{*}/3$, $p_{v}^{*}=\varepsilon_{v}^{*}/3$, $p^{*}=\varepsilon^{*}/3$, $p_{E}^{c}=\varepsilon_{E}^{c}/3$, $p_{v}^{c}=\varepsilon_{v}^{c}/3$ and $p^{c}=\varepsilon^{c}/3$. Moreover, for $m=0$ the trace anomaly (\ref{tt1}) is equal to zero, $\Delta_{2}^{c}=0$.

The Riemann integral of a real-valued function $f(x)$ defined on the interval $[a,b]$ is the limit of the Riemann sum:
\begin{equation}\label{rr1}
  \int\limits_{a}^{b} f(x) dx = \lim_{\lambda_{R}\to 0} \sum\limits_{k=1}^{n} f(\xi_{k}) \Delta x_{k},
\end{equation}
where a partition of an interval $[a,b]$ is a finite sequence of numbers of the form $a=x_{0}<x_{1}<\cdots<x_{n-1}<x_{n}=b$ with $n$ subintervals $[x_{k-1},x_{k}]$ of the length  $\Delta x_{k}=x_{k}-x_{k-1}$ $(\Delta x_{k}>0)$, $\lambda_{R}$ is the length of the longest subinterval and $x_{k-1}\leq\xi_{k}\leq x_{k}$. Let us rewrite Eqs.~(\ref{59}), (\ref{60}) in the form
\begin{eqnarray}\label{rr2}
  \varepsilon_{v}^{c} &=& \frac{1}{2(2\pi)^{3}} \sum\limits_{k_{x}=-\infty}^{\infty} \sum\limits_{k_{y}=-\infty}^{\infty} \sum\limits_{k_{z}=-\infty}^{\infty}
  \sqrt{p_{x}^{2}+p_{y}^{2}+p_{z}^{2}+m^{2}} \ \Delta p_{x} \Delta p_{y} \Delta p_{z}, \\ \label{rr3}
  \varepsilon^{c} &=& \frac{1}{(2\pi)^{3}} \sum\limits_{k_{x}=-\infty}^{\infty} \sum\limits_{k_{y}=-\infty}^{\infty} \sum\limits_{k_{z}=-\infty}^{\infty}
  \frac{\sqrt{p_{x}^{2}+p_{y}^{2}+p_{z}^{2}+m^{2}}}{e^{\beta\sqrt{p_{x}^{2}+p_{y}^{2}+p_{z}^{2}+m^{2}}}-1} \ \Delta p_{x} \Delta p_{y} \Delta p_{z},
\end{eqnarray}
where $p_{\alpha}=\Delta p_{\alpha} k_{\alpha}$ $(\alpha=x,y,z)$ with the length of subintervals $\Delta p_{\alpha}=\Delta p=2\pi/V^{1/3}$. In the thermodynamic limit as $V\to\infty$ we have $\Delta p \to 0$. Using Eq.~(\ref{rr1}), we obtain
\begin{eqnarray}\label{rr4}
  \varepsilon_{v}^{c} &=& \frac{1}{2(2\pi)^{3}} \lim_{\Delta p \to 0} \sum\limits_{k_{x}=-\infty}^{\infty} \sum\limits_{k_{y}=-\infty}^{\infty} \sum\limits_{k_{z}=-\infty}^{\infty}
  \sqrt{p_{x}^{2}+p_{y}^{2}+p_{z}^{2}+m^{2}} \ \Delta p_{x} \Delta p_{y} \Delta p_{z} \nonumber \\
  &=& \frac{1}{2(2\pi)^{3}} \int\limits_{-\infty}^{\infty} \int\limits_{-\infty}^{\infty}\int\limits_{-\infty}^{\infty} \sqrt{p_{x}^{2}+p_{y}^{2}+p_{z}^{2}+m^{2}} \ dp_{x} dp_{y} dp_{z}
\end{eqnarray}
and
\begin{eqnarray}\label{rr5}
  \varepsilon^{c} &=& \frac{1}{(2\pi)^{3}} \lim_{\Delta p \to 0} \sum\limits_{k_{x}=-\infty}^{\infty} \sum\limits_{k_{y}=-\infty}^{\infty} \sum\limits_{k_{z}=-\infty}^{\infty}
  \frac{\sqrt{p_{x}^{2}+p_{y}^{2}+p_{z}^{2}+m^{2}}}{e^{\beta\sqrt{p_{x}^{2}+p_{y}^{2}+p_{z}^{2}+m^{2}}}-1} \ \Delta p_{x} \Delta p_{y} \Delta p_{z} \nonumber \\
  &=& \frac{1}{(2\pi)^{3}} \int\limits_{-\infty}^{\infty} \int\limits_{-\infty}^{\infty} \int\limits_{-\infty}^{\infty} \frac{\sqrt{p_{x}^{2}+p_{y}^{2}+p_{z}^{2}+m^{2}}}{e^{\beta\sqrt{p_{x}^{2}+p_{y}^{2}+p_{z}^{2}+m^{2}}}-1} \
  dp_{x} dp_{y} dp_{z},
\end{eqnarray}
where, now, the momentum components $p_{x},p_{y}$ and $p_{z}$ under the integrals are continuum quantities. Then, for $m=0$ in the thermodynamic limit as $V\to\infty$ $(\Delta p \to 0)$, we obtain
\begin{eqnarray}\label{rr6}
  \varepsilon_{v}^{c} &=& \frac{1}{4\pi^{2}} \int\limits_{0}^{\infty} p^{3} dp  \to \infty, \\ \label{rr7}
  \varepsilon^{c} &=& \frac{1}{2\pi^{2}} \int\limits_{0}^{\infty} \frac{p^{3} dp}{e^{\beta p}-1}=\frac{\pi^{2}}{30\beta^{4}}=\varepsilon_{SB}.
\end{eqnarray}
These are values of the vacuum and physical terms of the energy density (\ref{58}) in the Stefan-Boltzmann limit. Since for the massless neutral scalar field the pressure $p_{v}^{c}=\varepsilon_{v}^{c}/3$ and $p^{c}=\varepsilon^{c}/3$, we have $p_{v}^{c} \to \infty$ and $p^{c}=\pi^{2}/(90\beta^{4})=p_{SB}$. The vacuum term of the density of the thermodynamic potential in the Stefan-Boltzmann limit is also divergent, $\omega_{v}^{c}=\varepsilon_{v}^{c}\to\infty$. However, the physical term of the density of the thermodynamic potential is
\begin{equation}\label{rr8}
  \omega_{ph}^{c} = \frac{1}{2\pi^{2}\beta} \int\limits_{0}^{\infty} p^{2} \ln\left( 1- e^{-\beta p} \right) dp = -\frac{\pi^{2}}{90\beta^{4}}=\omega_{SB}.
\end{equation}
Thus, the Stefan-Boltzmann limit corresponds to the continuum quantities of the massless neutral scalar field given in the thermodynamic limit, when the discrete momentum spectrum of bosons becomes continuous.

It should be stressed that at finite values of volume $V$ the shift $\Delta_{SB}=\varepsilon_{SB}-\varepsilon^{c}$ for the massless neutral scalar field is not equal to zero and the energy density $\varepsilon^{c}$ deviates from its Stefan-Boltzmann limit. We have
\begin{eqnarray}\label{rr9}
  \Delta_{SB} &=& \frac{1}{(2\pi)^{3}} \int\limits_{-\infty}^{\infty} \int\limits_{-\infty}^{\infty} \int\limits_{-\infty}^{\infty} \frac{\sqrt{p_{x}^{2}+p_{y}^{2}+p_{z}^{2}}}{e^{\beta\sqrt{p_{x}^{2}+p_{y}^{2}+p_{z}^{2}}}-1} \ dp_{x} dp_{y} dp_{z} - \frac{1}{V} \sum\limits_{k_{x}=-\infty}^{\infty} \sum\limits_{k_{y}=-\infty}^{\infty} \sum\limits_{k_{z}=-\infty}^{\infty}
  \frac{\sqrt{(\Delta p)^{2}(k_{x}^{2}+k_{y}^{2}+k_{z}^{2})}}{e^{\beta\sqrt{(\Delta p)^{2}(k_{x}^{2}+k_{y}^{2}+k_{z}^{2})}}-1}.
\end{eqnarray}
For the numerical proof of this statement see Section~\ref{sec5}.

\section{Physical thermodynamic quantities on a finite lattice}~\label{sec4}
We define the vacuum and the physical terms of thermodynamic quantities on the finite lattice $(N_{\beta},N_{\sigma})$ as in Ref.~\cite{Engels}. In this case, the vacuum terms of the thermodynamic quantities on the finite lattice are derived from Eqs.~(\ref{37})--(\ref{39}) by changing the summation over $k_{\beta}$ to integration, which corresponds to $T=0$. Thus, by definition, we have the lattice vacuum terms as~\cite{Engels}
\begin{eqnarray}\label{68}
  \omega_{v} &=& \frac{1}{a_{\beta} V} \sum\limits_{\vec{k}=k_{\sigma 1}}^{k_{\sigma 2}} \ln\left[\frac{a_{\beta}\omega}{2}+ \left[1+\left(\frac{a_{\beta}\omega}{2}\right)^{2}\right]^{1/2}\right],  \\ \label{69}
   \varepsilon_{v}  &=&  \frac{1}{2V} \sum\limits_{\vec{k}=k_{\sigma 1}}^{k_{\sigma 2}} \omega  \left[1+\left(\frac{a_{\beta}\omega}{2}\right)^{2}\right]^{-1/2}, \\ \label{70}
   p_{v} &=& \frac{1}{6V} \sum\limits_{\vec{k}=k_{\sigma 1}}^{k_{\sigma 2}} \left(\omega-\frac{m^{2}}{\omega}\right) \left[1+\left(\frac{a_{\beta}\omega}{2}\right)^{2}\right]^{-1/2},
\end{eqnarray}
where $a_{\beta}=const$ and $\omega$ is given in Eq.~(\ref{36}). Then, the physical terms of the thermodynamic quantities on the finite lattice can be written as~\cite{Engels}
\begin{eqnarray}\label{71}
 \omega_{ph} &=& \omega_{E} - \omega_{v}= \frac{1}{2\beta V}\sum\limits_{\vec{k}=k_{\sigma 1}}^{k_{\sigma 2}}\sum\limits_{k_{\beta}=k_{\beta 1}}^{k_{\beta 2}} \ln\left[a_{\beta}^{2}(\omega_{\beta}^{2} + \omega^{2})\right] - \frac{1}{a_{\beta} V} \sum\limits_{\vec{k}=k_{\sigma 1}}^{k_{\sigma 2}} \ln\left[\frac{a_{\beta}\omega}{2}+ \left[1+\left(\frac{a_{\beta}\omega}{2}\right)^{2}\right]^{1/2}\right], \\ \label{72}
 \varepsilon &=& \varepsilon_{E}-\varepsilon_{v} = \frac{1}{\beta V}\sum\limits_{\vec{k}=k_{\sigma 1}}^{k_{\sigma 2}} \sum\limits_{k_{\beta}=k_{\beta 1}}^{k_{\beta 2}} \frac{\omega^{2}}{\omega_{\beta}^{2} + \omega^{2}} - \frac{1}{2V} \sum\limits_{\vec{k}=k_{\sigma 1}}^{k_{\sigma 2}} \omega  \left[1+\left(\frac{a_{\beta}\omega}{2}\right)^{2}\right]^{-1/2}, \\ \label{73}
   p &=& p_{E} - p_{v}= \frac{1}{3\beta V}\sum\limits_{\vec{k}=k_{\sigma 1}}^{k_{\sigma 2}} \sum\limits_{k_{\beta}=k_{\beta 1}}^{k_{\beta 2}} \frac{\omega^{2}-m^{2}}{\omega_{\beta}^{2} + \omega^{2}} - \frac{1}{6V} \sum\limits_{\vec{k}=k_{\sigma 1}}^{k_{\sigma 2}} \left(\omega-\frac{m^{2}}{\omega}\right) \left[1+\left(\frac{a_{\beta}\omega}{2}\right)^{2}\right]^{-1/2}.
\end{eqnarray}
These quantities are the same as those from Ref.~\cite{Engels}. Here and below for $m=0$ the $\vec{k}=0,k_{\beta}=0$ term does not contribute to the sum.

Using Eqs.~(\ref{72}) and (\ref{73}), we obtain the trace anomaly for the free real scalar field on the finite lattice as
\begin{equation}\label{tt3}
  \Delta_{2} = \beta^{4} (\varepsilon-3p)= \frac{\beta^{3}}{V}\sum\limits_{\vec{k}=k_{\sigma 1}}^{k_{\sigma 2}}\sum\limits_{k_{\beta}=k_{\beta 1}}^{k_{\beta 2}}  \frac{m^{2}}{\omega_{\beta}^{2} + \omega^{2}} - \frac{\beta^{4}}{2V} \sum\limits_{\vec{k}=k_{\sigma 1}}^{k_{\sigma 2}} \frac{m^{2}}{\omega} \left[1+\left(\frac{a_{\beta}\omega}{2}\right)^{2}\right]^{-1/2}.
\end{equation}
It defines the deviation of the equation of state from the Stefan-Boltzmann equation, $p_{SB}=\varepsilon_{SB}/3$. The trace anomaly for the free massive real scalar field on the finite lattice is not equal to zero at the finite temperature and volume. However, for $m= 0$ the trace anomaly (\ref{tt3}) is equal to zero,  $\Delta_{2}=0$.

The potential inhomogeneity for the free real scalar field on the finite lattice is given by
\begin{equation}\label{pp3}
  \Delta_{1} = 3\beta^{4} (\omega_{ph}+p).
\end{equation}
Its finite values define the violation of the homogeneous properties of the thermodynamic potential on a finite lattice $(N_{\beta},N_{\sigma})$. See Section~\ref{ap1}.

It should be stressed that in the continuum limit the vacuum and physical lattice quantities given in Eqs.~(\ref{68}), (\ref{69}), (\ref{70}), (\ref{71}), (\ref{72}) and (\ref{73}) resemble the continuum limit quantities (\ref{66}), (\ref{59}), (\ref{63}), (\ref{67}), (\ref{60}) and (\ref{64}), respectively. In the continuum limit we have
$\omega=\sqrt{\vec{p}^{2}+m^{2}}$, $p_{\alpha}=\Delta p_{\alpha} k_{\alpha}$ $(\alpha=x,y,z)$, $k_{\sigma 1}=-\infty$ and $k_{\sigma 2}=\infty$. Substituting these quantities in Eqs.~(\ref{68}), (\ref{69}), (\ref{70}) and taking the limit $a_{\beta}\to 0$, we obtain
\begin{eqnarray}\label{68a}
  \omega_{v} &=& \frac{1}{V} \lim_{a_{\beta}\to 0} \sum\limits_{\vec{k}}  \ln\left[\frac{a_{\beta}\omega}{2}+ \left[1+\left(\frac{a_{\beta}\omega}{2}\right)^{2}\right]^{1/2}\right]^{\frac{1}{a_{\beta}}} = \frac{1}{V}  \sum\limits_{\vec{k}} \ln e^{\frac{\omega}{2}} =
  \frac{1}{2V} \sum\limits_{\vec{k}} \omega =\omega_{v}^{c},  \\ \label{69a}
   \varepsilon_{v}  &=& \frac{1}{2V} \lim_{a_{\beta}\to 0} \sum\limits_{\vec{k}}  \omega  \left[1+\left(\frac{a_{\beta}\omega}{2}\right)^{2}\right]^{-1/2}= \frac{1}{2V} \sum\limits_{\vec{k}} \omega = \varepsilon_{v}^{c} , \\ \label{70a}
   p_{v} &=& \frac{1}{6V} \lim_{a_{\beta}\to 0} \sum\limits_{\vec{k}} \left(\omega-\frac{m^{2}}{\omega}\right) \left[1+\left(\frac{a_{\beta}\omega}{2}\right)^{2}\right]^{-1/2} =\frac{1}{6V} \sum\limits_{\vec{k}}  \left(\omega-\frac{m^{2}}{\omega}\right) = p_{v}^{c},
\end{eqnarray}
where the range of $k_{\alpha}$ is given below Eq.~(\ref{60}). In the previous Section it was proved that in the continuum limit we have $\omega_{E}=\omega_{E}^{c}$, $\varepsilon_{E}=\varepsilon_{E}^{c}$ and $p_{E}=p_{E}^{c}$. Then, using Eqs.~(\ref{65}), (\ref{58}), (\ref{62}) and Eqs.~(\ref{71}), (\ref{72}), (\ref{73}), we obtain $\omega_{ph} = \omega_{E}^{c} - \omega_{v}^{c}=\omega_{ph}^{c}$, $\varepsilon = \varepsilon_{E}^{c} - \varepsilon_{v}^{c}=\varepsilon^{c}$ and $p = p_{E}^{c} - p_{v}^{c}=p^{c}$. Thus, in the continuum limit the vacuum and physical terms of the thermodynamic quantities given in Ref.~\cite{Engels} exactly coincide with the same quantities obtained in the canonical quantization method~\cite{Kapusta89,Bellac}.

In the limit $N_{\beta}\to\infty,a_{\beta}\to 0$ at $\beta=const$ and fixed value of $N_{\sigma}$ the vacuum and physical lattice quantities given in Eqs.~(\ref{68}), (\ref{69}), (\ref{70}), (\ref{71}), (\ref{72}) and (\ref{73}) recover Eqs.~(\ref{56}), (\ref{45}), (\ref{50}), (\ref{57}), (\ref{46}) and (\ref{51}), respectively. In this limit Eqs.~(\ref{68}), (\ref{69}), (\ref{70}) can be rewritten as
\begin{eqnarray}\label{68b}
  \omega_{v} &=& \frac{1}{V} \lim_{a_{\beta}\to 0} \sum\limits_{\vec{k}=k_{\sigma 1}}^{k_{\sigma 2}}  \ln\left[\frac{a_{\beta}\omega}{2}+ \left[1+\left(\frac{a_{\beta}\omega}{2}\right)^{2}\right]^{1/2}\right]^{\frac{1}{a_{\beta}}} = \frac{1}{V}  \sum\limits_{\vec{k}=k_{\sigma 1}}^{k_{\sigma 2}} \ln e^{\frac{\omega}{2}} = \frac{1}{2V} \sum\limits_{\vec{k}=k_{\sigma 1}}^{k_{\sigma 2}} \omega =\omega_{v}^{*},  \\ \label{69b}
   \varepsilon_{v}  &=& \frac{1}{2V} \lim_{a_{\beta}\to 0} \sum\limits_{\vec{k}=k_{\sigma 1}}^{k_{\sigma 2}}  \omega  \left[1+\left(\frac{a_{\beta}\omega}{2}\right)^{2}\right]^{-1/2}=  \frac{1}{2V} \sum\limits_{\vec{k}=k_{\sigma 1}}^{k_{\sigma 2}} \omega = \varepsilon_{v}^{*} , \\ \label{70b}
   p_{v} &=& \frac{1}{6V} \lim_{a_{\beta}\to 0} \sum\limits_{\vec{k}=k_{\sigma 1}}^{k_{\sigma 2}} \left(\omega-\frac{m^{2}}{\omega}\right) \left[1+\left(\frac{a_{\beta}\omega}{2}\right)^{2}\right]^{-1/2} = \frac{1}{6V} \sum\limits_{\vec{k}=k_{\sigma 1}}^{k_{\sigma 2}}  \left(\omega-\frac{m^{2}}{\omega}\right) = p_{v}^{*},
\end{eqnarray}
where $\omega$ is given in Eq.~(\ref{36}). In the previous Section it was proved that in this limit we have $\omega_{E}=\omega_{E}^{*}$, $\varepsilon_{E}=\varepsilon_{E}^{*}$ and $p_{E}=p_{E}^{*}$. Then, using Eqs.~(\ref{55}), (\ref{44}), (\ref{49}) and Eqs.~(\ref{71}), (\ref{72}), (\ref{73}), we obtain $\omega_{ph} = \omega_{E}^{*} - \omega_{v}^{*}=\omega_{ph}^{*}$, $\varepsilon = \varepsilon_{E}^{*} - \varepsilon_{v}^{*}=\varepsilon^{*}$ and $p = p_{E}^{*} - p_{v}^{*}=p^{*}$.

\section{Thermodynamic potential and zeroth law of thermodynamics}\label{ap1}
In the equilibrium statistical mechanics all thermodynamic quantities belong to the class of homogeneous functions of the zero and first order. This property of the thermodynamic quantities provides the fulfilment of the requirements of the equilibrium thermodynamics. For example, in the grand canonical ensemble the zeroth law of thermodynamics is satisfied if the temperature is intensive (zero order) and the grand thermodynamic potential is extensive, i.e. it is a homogeneous function of the first order with respect to the extensive variable of state $V$.

Let us consider the case when the thermodynamic potential of the grand canonical ensemble is indeed a homogeneous function of the first order with respect to the extensive variable of state $V$. We have~\cite{Parvan2015,Parvan2015a}
\begin{equation}\label{y1}
      \Omega(T,V,\mu) = V \omega(T,\mu).
\end{equation}
Then the pressure $p$ and the potential inhomogeneity $\Delta_{1}=3(\omega+p)/T^{4}$ can be written as
\begin{equation}\label{y2}
  p = -\frac{\partial \Omega}{\partial V} =-\frac{\Omega}{V}
\end{equation}
and
\begin{equation}\label{y3}
  \Delta_{1} = \frac{3}{T^{4}}\left(\frac{\Omega}{V}+p\right)=0.
\end{equation}
For the free neutral scalar field $\mu=0$. Note that only in this section $\omega$ denotes the density of the thermodynamic potential of the grand canonical ensemble.

To prove the zeroth law of thermodynamics, let us divide the system into two subsystems ($1$ and $2$). Then the extensive variables of state of the grand canonical ensemble should be additive and the intensive variables of state should be the same~\cite{Parvan2015,Parvan2015a}
\begin{equation}\label{y4}
  V=V^{1}+V^{2}, \quad T=T^{1}=T^{2}, \quad \mu=\mu^{1}=\mu^{2}.
\end{equation}
Since the function $\omega$ in Eq.~(\ref{y1}) depends only on the intensive variables of state $T$ and $\mu$ then we have
\begin{equation}\label{y5}
  \omega(T,\mu)=\omega^{1}(T^{1},\mu^{1})=\omega^{2}(T^{2},\mu^{2}).
\end{equation}
Using Eqs.~(\ref{y1}), (\ref{y4}) and (\ref{y5}), we obtain
\begin{equation}\label{y6}
 \Omega^{1}(T^{1},V^{1},\mu^{1})+\Omega^{2}(T^{2},V^{2},\mu^{2})=\Omega(T,V,\mu).
\end{equation}
Thus, we have obtained that if the thermodynamic potential is a homogeneous function of the first order and the temperature is intensive, then the thermodynamic potential is an additive function and the potential inhomogeneity is zero. This proves the zeroth law of thermodynamics for the grand canonical ensemble~\cite{Parvan2015,Parvan2015a}.

Let us consider a more general case when the thermodynamic potential of the grand canonical ensemble is an inhomogeneous function. For example, we can write
\begin{equation}\label{y7}
      \Omega(T,V,\mu) = V^{\alpha} \tilde{\omega}(T,\mu),
\end{equation}
where $\alpha$ is a real number. For $\alpha=1$ we have Eq.~(\ref{y1}). Then the pressure $p$ and the potential inhomogeneity $\Delta_{1}$ can be written as
\begin{equation}\label{y8}
  p = -\frac{\partial \Omega}{\partial V} =-\alpha \frac{\Omega}{V}
\end{equation}
and
\begin{equation}\label{y9}
  \Delta_{1} = \frac{3}{T^{4}}\left(\frac{\Omega}{V}+p\right)=\frac{3}{T^{4}}(1-\alpha)\frac{\Omega}{V}.
\end{equation}
For the inhomogeneous thermodynamic potential (\ref{y7}) at $\alpha\neq 1$ the potential inhomogeneity $\Delta_{1}\neq 0$ and the pressure $p\neq -\Omega/V$.

To verify the zeroth law of thermodynamics for the inhomogeneous thermodynamic potential (\ref{y7}), let us divide the system into two subsystems (1 and 2) and require Eq.~(\ref{y4}). Then we have
\begin{equation}\label{y11}
  \tilde{\omega}(T,\mu) = \tilde{\omega}^{1}(T^{1},\mu^{1}) = \tilde{\omega}^{2}(T^{2},\mu^{2}).
\end{equation}
Using Eqs.~(\ref{y4}), (\ref{y7}) and (\ref{y11}), we obtain
\begin{equation}\label{y12}
 \Omega^{1}(T^{1},V^{1},\mu^{1}) + \Omega^{2}(T^{2},V^{2},\mu^{2}) = \frac{(V^{1})^{\alpha}+(V^{2})^{\alpha}}{V^{\alpha}}\ \Omega(T,V,\mu). \;\;\;\;\;
\end{equation}
For $\alpha\neq 1$ we have $\Omega^{1}+\Omega^{2}\neq \Omega$. Thus, we have obtained that if the thermodynamic potential of the grand canonical ensemble is not a homogeneous function of the first order with respect to the extensive variable of state $V$, then the thermodynamic potential $\Omega$ is nonadditive, the potential inhomogeneity $\Delta_{1}$ is not equal to zero ($p\neq -\Omega/V$) and the zeroth law of thermodynamics is not satisfied.

\section{Analysis and Results}~\label{sec5}
Let us study numerically the thermodynamic properties and the finite volume corrections to the thermodynamic quantities for the real scalar field on the finite lattice and in the continuum limit at a finite temperature $T$ in some finite volume $V$ which is typical for the ultrarelativistic heavy ion and hadron-hadron collisions~\cite{Abelev14,Aamodt11,Khachatryan10_2,Mercado11,Mizoguchi10,Csorgo06,Adare10,Wilde13}.

\begin{figure}[tp]
\includegraphics[width=0.96\textwidth]{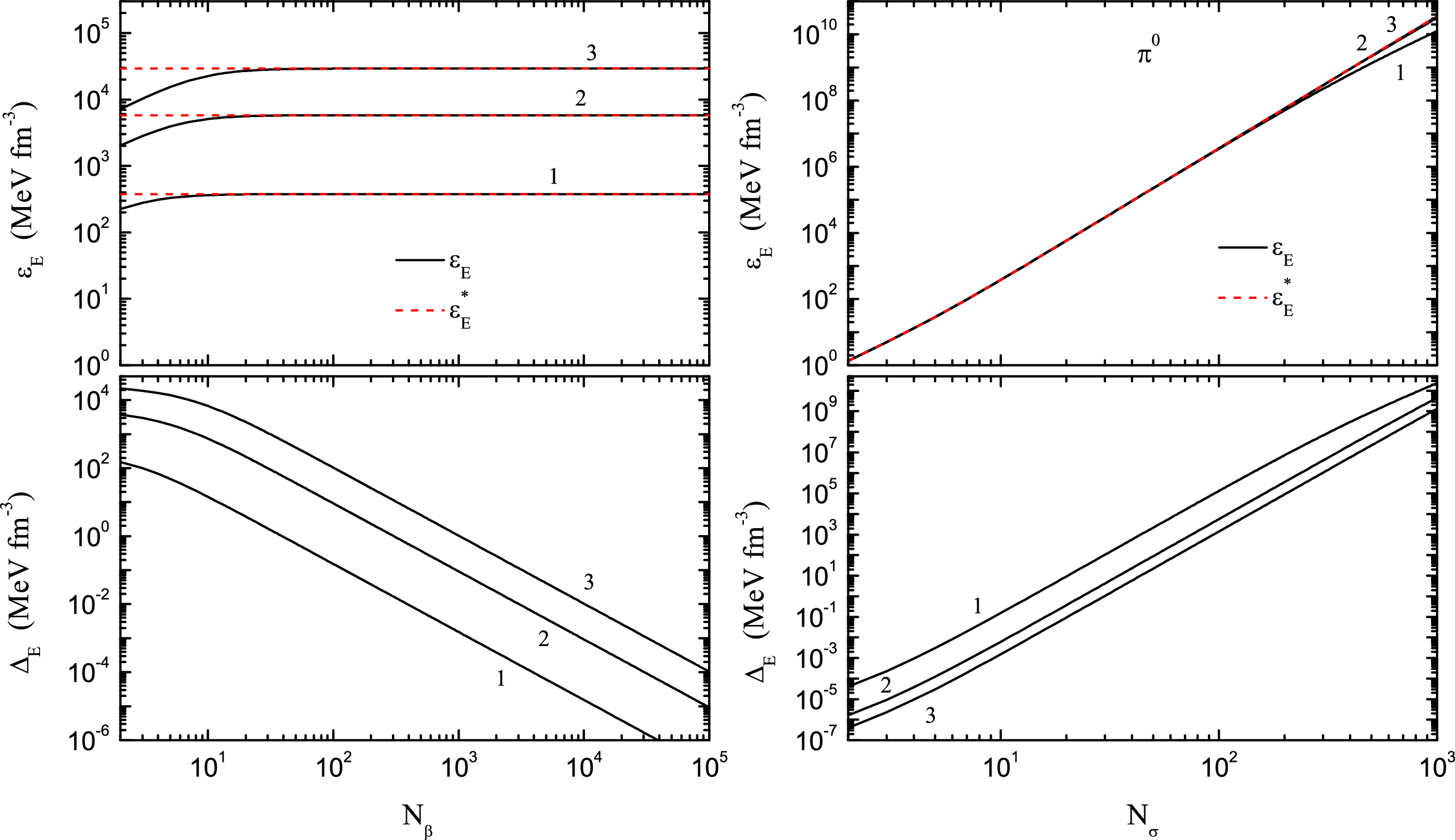}
\caption{(Color online) The energy density $\varepsilon_{E}$, its limit $\varepsilon_{E}^{*}$, and the shift  $\Delta_{E}=\varepsilon_{E}^{*}-\varepsilon_{E}$ as functions of $N_{\beta}$ and $N_{\sigma}$ for the free neutral scalar field on the finite lattice with the mass of $\pi^{0}$ pion at temperature $T=100$ MeV in the volume $V=9^{3}$ fm$^{3}$. Left panels: The lines $1,2$ and $3$ correspond to the values of $N_{\sigma}=10,20$ and $30$, respectively. Right panels: The lines $1,2$ and $3$ correspond to the values of $N_{\beta}=100,500$ and $1000$, respectively.} \label{fig1}
\end{figure}

Figure~\ref{fig1} shows the dependence of the energy density $\varepsilon_{E}$, its limit $\varepsilon_{E}^{*}$, and the shift $\Delta_{E}\equiv\varepsilon_{E}^{*}-\varepsilon_{E}$ on the number of sites $N_{\beta}$ and $N_{\sigma}$ for the free neutral scalar field on the finite lattice with the mass of $\pi^{0}$ pion at temperature $T$ in volume $V$. For the given $T$, $V$ and $N_{\sigma}$ (left panels), the energy density $\varepsilon_{E}$ (\ref{38}) increases with $N_{\beta}$ and attains the constant value $\varepsilon_{E}^{*}$ (\ref{44}) at large values of $N_{\beta}$. The shift $\Delta_{E}$ decreases with increasing $N_{\beta}$ and tends to zero, $\Delta_{E}=0$, as $N_{\beta}\to\infty$. For the given $T$, $V$ and $N_{\beta}$ (right panels), the energy density $\varepsilon_{E}$, its limit $\varepsilon_{E}^{*}$ and the shift $\Delta_{E}$ increase with $N_{\sigma}$. However, at large values of $N_{\beta}$ the energy density $\varepsilon_{E}$ tends to its limiting value $\varepsilon_{E}^{*}$ and the shift $\Delta_{E}$ decreases with increasing $N_{\beta}$.

\begin{figure}
\includegraphics[width=0.96\textwidth]{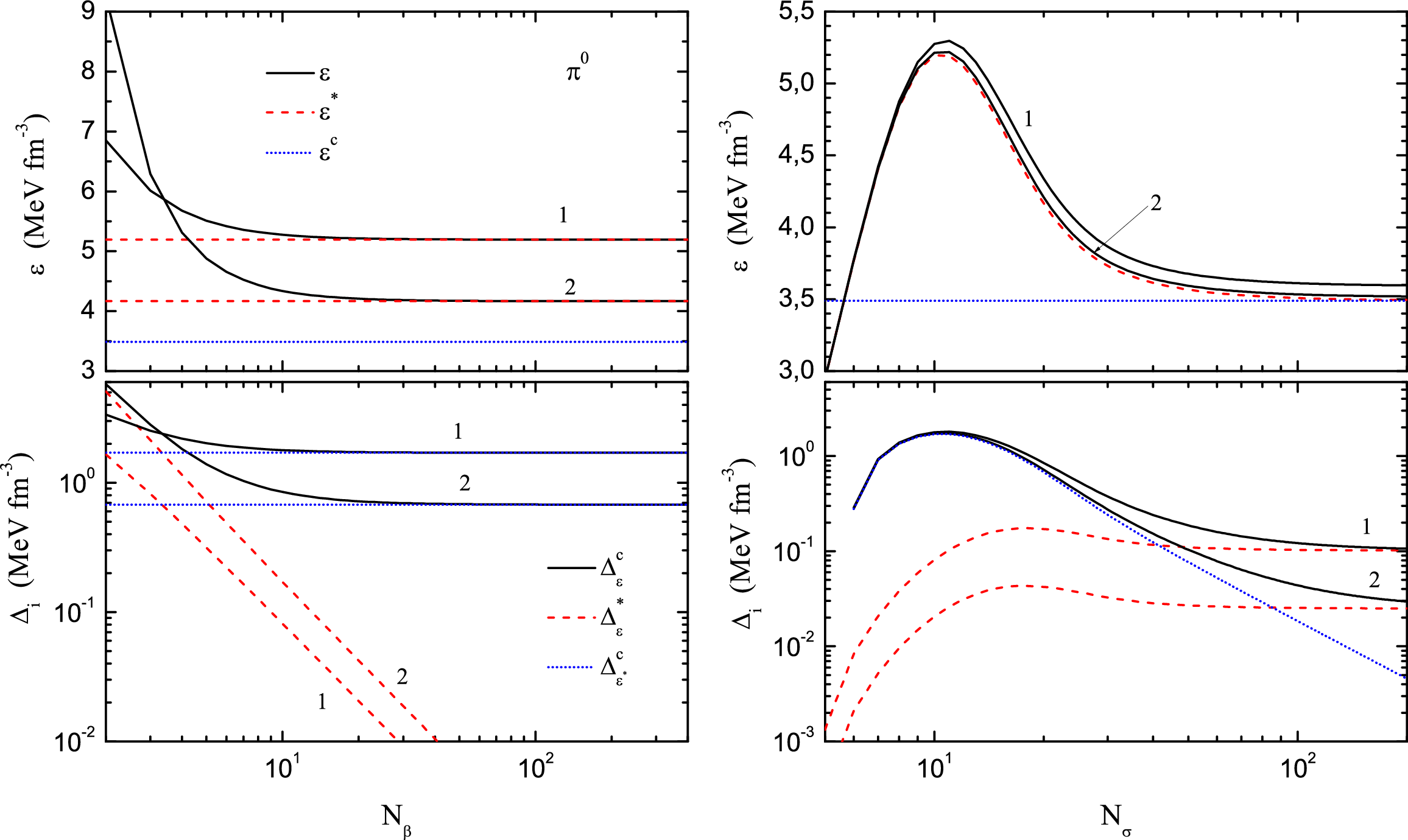}
\caption{(Color online) The physical energy density $\varepsilon$, its limit $\varepsilon^{*}$, and continuum limit $\varepsilon^{c}$, and the shifts $\Delta_{\varepsilon}^{c}=\varepsilon-\varepsilon^{c}$, $\Delta_{\varepsilon}^{*}=\varepsilon-\varepsilon^{*}$, $\Delta_{\varepsilon^{*}}^{c}=\varepsilon^{*}-\varepsilon^{c}$ as functions of $N_{\beta}$ and $N_{\sigma}$ for the free neutral scalar field on the finite lattice with the mass of $\pi^{0}$ pion at a temperature $T=100$ MeV in the volume $V=9^{3}$ fm$^{3}$. Left panels: The lines $1$ and $2$ correspond to the values of $N_{\sigma}=10$ and $20$, respectively. Right panels: The lines $1$ and $2$ correspond to the values of $N_{\beta}=10$ and $20$, respectively.} \label{fig2}
\end{figure}

Figure~\ref{fig2} represents the dependence of the physical energy density $\varepsilon$, its limits $\varepsilon^{*}$ and $\varepsilon^{c}$, and the shifts
$\Delta_{\varepsilon}^{c}\equiv\varepsilon-\varepsilon^{c}$, $\Delta_{\varepsilon}^{*}\equiv\varepsilon-\varepsilon^{*}$, and $\Delta_{\varepsilon^{*}}^{c}\equiv\varepsilon^{*}-\varepsilon^{c}$ on the lattice parameters $N_{\beta}$ and $N_{\sigma}$ for the free neutral scalar field on the finite lattice with the mass of $\pi^{0}$ pion at some fixed temperature and volume. For the given $T$, $V$ and fixed $N_{\sigma}$ (left panels), the physical energy density $\varepsilon$ (\ref{72}) decreases with $N_{\beta}$ and tends to the constant $\varepsilon^{*}$ (\ref{46}) as $N_{\beta}\to\infty$, and the shift $\Delta_{\varepsilon}^{*}$ goes to zero. However, $\Delta_{\varepsilon}^{c}$ approaches $\Delta_{\varepsilon^{*}}^{c}$ and they become the same constants different from zero as $N_{\beta}\to\infty$. For large values of $N_{\beta}$ the physical energy densities $\varepsilon$ and $\varepsilon^{*}$ decrease with increasing $N_{\sigma}$ and tend to their limiting value $\varepsilon^{c}$ (\ref{60}), and the shifts $\Delta_{\varepsilon}^{c}$ and $\Delta_{\varepsilon^{*}}^{c}$ go to zero. For the given $T$, $V$ and fixed $N_{\beta}$ (right panels), the physical energy density $\varepsilon$
has a maximum and then decreases with increasing $N_{\sigma}$, and tends to a constant value different from $\varepsilon^{*}$ and $\varepsilon^{c}$. The lattice physical energy density $\varepsilon^{*}$ also has a maximum and decreases with increasing $N_{\sigma}$, but tends to the constant value $\varepsilon^{c}$ as $N_{\sigma}\to\infty$. The shift $\Delta_{\varepsilon^{*}}^{c}$ goes to zero with increasing $N_{\sigma}$; however, $\Delta_{\varepsilon}^{c}$ approaches $\Delta_{\varepsilon}^{*}$, and they become the same constants different from zero with increasing $N_{\sigma}$ for the fixed $N_{\beta}$; $\Delta_{\varepsilon}^{c}$ and $\Delta_{\varepsilon}^{*}$ tend to zero as $N_{\beta}\to\infty$ and $N_{\sigma}\to\infty$.

The exact calculations of the lattice physical energy densities (\ref{46}), (\ref{60}) and (\ref{72}) allow us to estimate the lattice size $(N_{\beta},N_{\sigma})$
which approximates the continuum limit with certain accuracy. The specific lattice parameter $N_{\sigma}$ can be found from the condition $\Delta_{\varepsilon^{*}}^{c}=\varepsilon^{*}-\varepsilon^{c}\leq \delta_{\sigma}$ with some given accuracy $\delta_{\sigma}$. Then, the specific lattice parameter $N_{\beta}$ can be estimated from the behavior of the  physical energy densities $\varepsilon$ and $\varepsilon^{*}$ as functions of $N_{\beta}$ at fixed $N_{\sigma}$, i.e., from the condition $\Delta_{\varepsilon}^{*}=\varepsilon-\varepsilon^{*}\leq 10^{-2} \delta_{\sigma}$. Here, the number $10^{-2}$ was chosen arbitrarily. 

\begin{figure}
\includegraphics[width=0.46\textwidth]{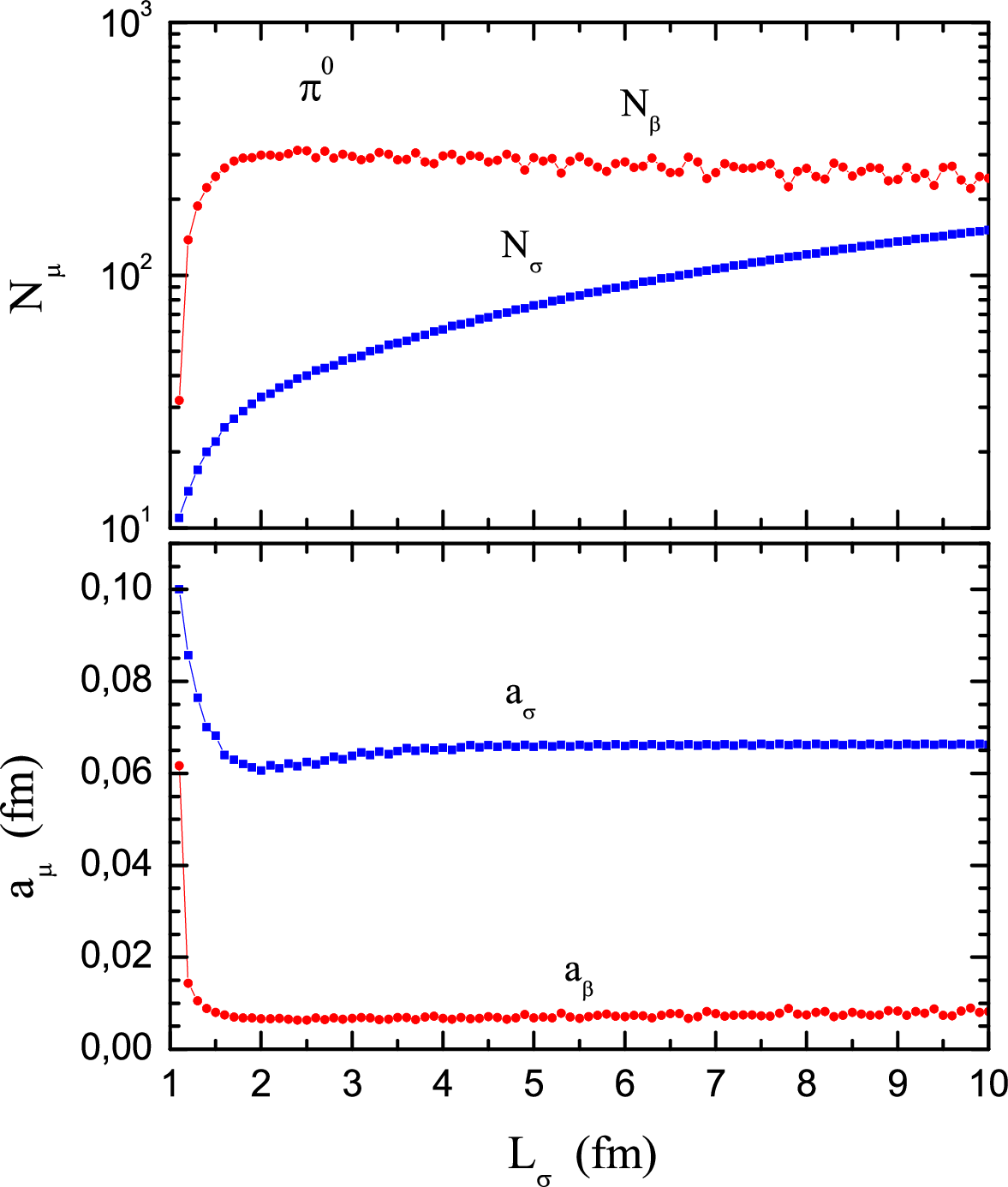}
\caption{(Color online) The values of the lattice parameters $N_{\beta}$, $N_{\sigma}$, $a_{\beta}$ and $a_{\sigma}$ which approximate the continuum limit of the physical energy density as functions of the length $L_{\sigma}$ for the free neutral scalar field on the finite lattice with the mass of $\pi^{0}$ pion at the temperature $T=100$ MeV.} \label{fig3}
\end{figure}

As an example, Figure~\ref{fig3} represents the number of lattice sites $N_{\beta}$, $N_{\sigma}$ and the spacings $a_{\beta}$, $a_{\sigma}$ which approximate the continuum limit for the physical energy density as functions of the length $L_{\sigma}$ for the free neutral scalar field on the finite lattice with the mass of $\pi^{0}$ pion at the temperature $T=100$ MeV and $\delta_{\sigma}=0.01$ MeV fm$^{-3}$. For the length $L_{\sigma}\leq 1.5$ fm, the numbers of lattice sites $N_{\beta}$, $N_{\sigma}$ grow rapidly with $L_{\sigma}$ and achieve the values $N_{\beta}\approx 300$ and $N_{\sigma}\approx 30$. Then, $N_{\sigma}$ continues to increase more slowly but $N_{\beta}$ slightly decreases. However, at large volumes this difference diminishes. For example, for $L_{\sigma}=10$ fm the parameter $N_{\sigma}=151$ and the parameter $N_{\beta}=242$. The spacings $a_{\beta}$, $a_{\sigma}$ are almost constant. For example, for $L_{\sigma}=10$ fm the parameter $a_{\sigma}=0.066$ fm and the parameter $a_{\beta}=0.008$ fm. Thus, with increasing volume the number of spatial sites $N_{\sigma}$, which gives a good approximation to the continuum limit, increases; however, the spatial spacing $a_{\sigma}$ may remain unchanged. The number of temporal sites $N_{\beta}$ and the temporal spacing $a_{\beta}$ are practically unchanged with $V$ at fixed values of the temperature $T$. Note that at these values of $T$, $V$, $N_{\beta}$ and $N_{\sigma}$ the energy density $\varepsilon_{E}$ deviates essentially from its limit $\varepsilon_{E}^{*}$. See Table~\ref{t1}.

\begin{figure}
\includegraphics[width=0.47\textwidth]{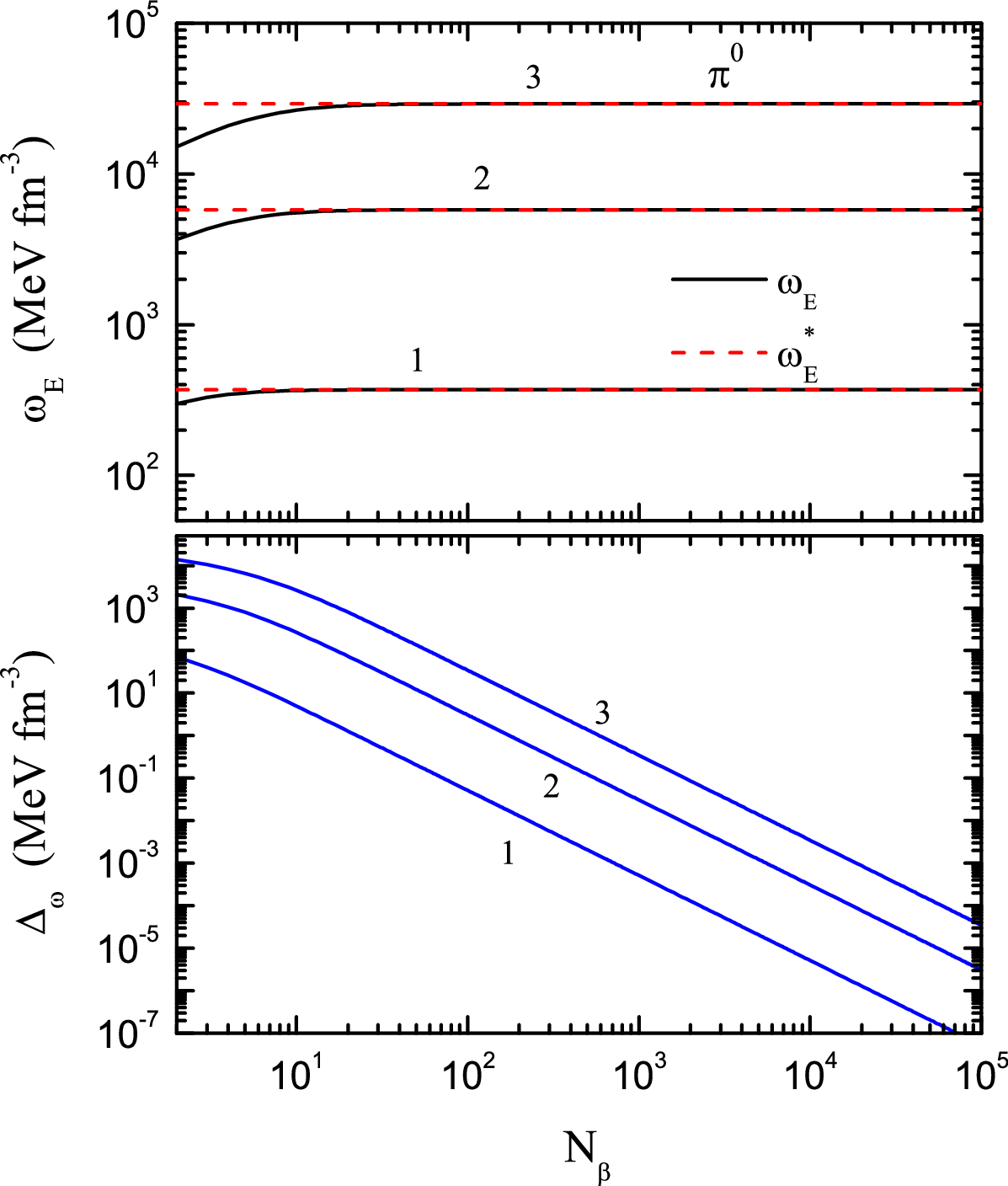}
\caption{(Color online) The density of the thermodynamic potential $\omega_{E}$ (\ref{37}), its limit $\omega_{E}^{*}$ (\ref{55}), and the shift $\Delta_{\omega}$ as functions of
$N_{\beta}$ for the free neutral scalar field on the finite lattice with the mass of $\pi^{0}$ pion at the temperature $T=100$ MeV in the volume $V=9^{3}$ fm$^{3}$ for different values of $N_{\sigma}$. The lines $1,2$ and $3$ correspond to the values of $N_{\sigma}=10,20$ and $30$, respectively.} \label{fig4}
\end{figure}

Let us numerically verify Eq.~(\ref{55}) noting that the density of the thermodynamic potential (\ref{37}) in the limit $N_{\beta}\to\infty$, $\beta=const$ and the fixed value of $N_{\sigma}$ is divided into the sum of two terms: the vacuum term and the physical term. Figure~\ref{fig4} represents the density of the thermodynamic potential $\omega_{E}$ (\ref{37}), its limit $\omega_{E}^{*}$ (\ref{55}), and the shift $\Delta_{\omega}\equiv \omega_{E}^{*}-\omega_{E}$ as functions of $N_{\beta}$ for the free neutral scalar field on the finite lattice with the mass of $\pi^{0}$ pion at the temperature $T$ in some volume $V$ for different values of $N_{\sigma}$. For the given $T$, $V$ and $N_{\sigma}$, the density of the thermodynamic potential $\omega_{E}$ (\ref{37}) increases with $N_{\beta}$ and attains the constant $\omega_{E}^{*}$ (\ref{55}) at large values of $N_{\beta}$. The shift $\Delta_{\omega}$ decreases with increasing $N_{\beta}$ and tends to zero, $\Delta_{\omega}=0$, as $N_{\beta}\to\infty$. This numerically proves that Eq.~(\ref{37}) in the limit $N_{\beta}\to\infty$, $\beta=const$ and fixed value of $N_{\sigma}$ recovers Eq.~(\ref{55}) which contains only two terms: the vacuum term $\omega_{v}^{*}$ and the physical term $\omega_{ph}^{*}$.

\begin{table}
  \caption{Some values of $\Delta_{E}=\varepsilon_{E}^{*}-\varepsilon_{E}$ for the free neutral scalar field on the finite lattice with the mass of $\pi^{0}$ pion at the temperature $T=100$ MeV for different values of $N_{\beta}$ and $V$. The shift $\Delta_{E}$ is given in units of MeV fm$^{-3}$.}\label{t1}
\begin{tabular}{ccccc}
  \hline
                            $V$ (fm$^{3}$)  & $ 1 $  & $ 3^{3}$  & $ 6^{3}$  &$ 9^{3} $   \\
  \hline
                                            & \multicolumn{4}{c}{$N_{\sigma}$}  \\
                     \cline{2-5}
   $N_{\beta}$  & $8$    & $47$      & $91$      &$136$   \\
  \hline
          $10^{3}$       &$1.97\;10^{2}$               &$1.11\;10^{4}$           &$9.14\;10^{3}$             &$ 8.94\;10^{3}$ \\
          $10^{4}$       &$1.97\;10^{0}$               &$1.11\;10^{2}$            &$9.15\;10^{1}$             &$8.95\;10^{1}$  \\
          $10^{5}$       &$1.97\;10^{-2}$              &$1.11\;10^{0}$            &$9.15\;10^{-1}$            &$8.97\;10^{-1}$  \\
\hline
\end{tabular}
\end{table}

\begin{figure}
\includegraphics[width=0.93\textwidth]{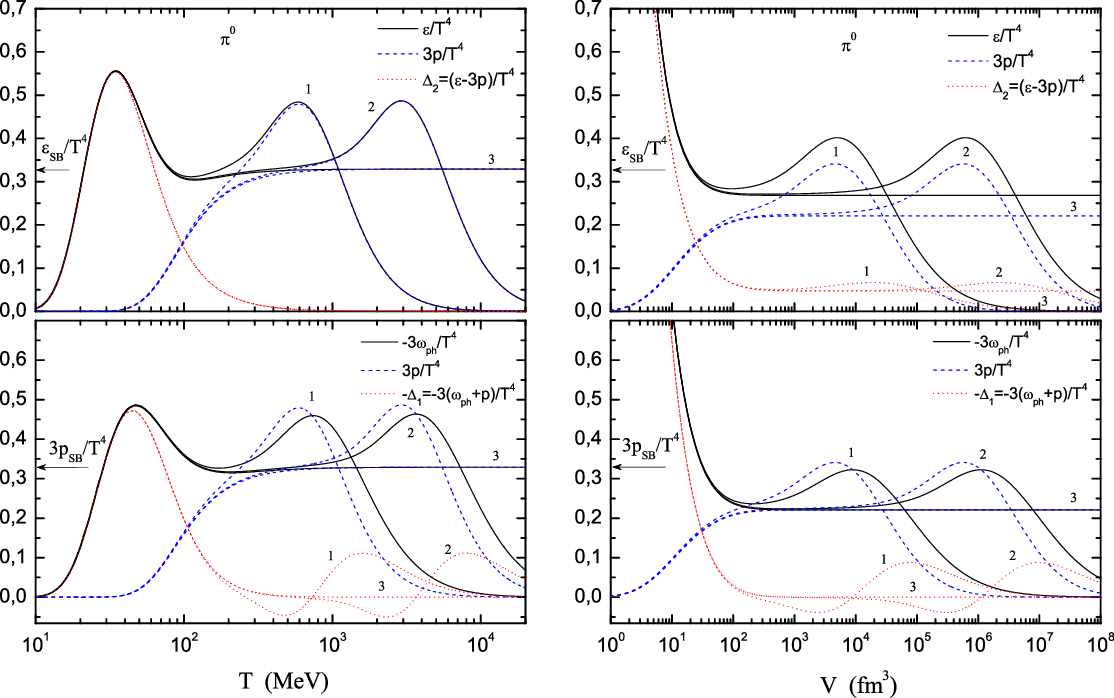}
\caption{(Color online) The exact lattice results for the massive scalar field. The physical energy density $\varepsilon$ (\ref{72}), the physical pressure $p$ (\ref{73}), the density of the physical thermodynamic potential $\omega_{ph}$ (\ref{71}), the trace anomaly $\Delta_{2}$ (\ref{tt3}) and the potential inhomogeneity $\Delta_{1}$ (\ref{pp3}) as functions of volume $V$ and temperature $T$ for the free neutral scalar field with the mass of $\pi^{0}$ pion on the lattice for $N_{\beta}=20$, $N_{\sigma}=20$ (line $1$), $N_{\sigma}=100$ (line $2$) and in the continuum limit (line $3$). Left panels: All curves were calculated at the volume $V=3^{3}$ fm$^{3}$. Right panels: All lines correspond to the temperature $T=100$ MeV.} \label{fig6}
\end{figure}

Figure~\ref{fig6} represents the behavior of the physical energy density (\ref{72}), the physical pressure (\ref{73}), the density of the physical thermodynamic potential (\ref{71}), the trace anomaly (\ref{tt3}) and the potential inhomogeneity (\ref{pp3}) as functions of volume $V$ and temperature $T$ for the free neutral scalar field on the finite lattice with the mass of $\pi^{0}$ pion. For $V=3^{3}$ fm$^{3}$ and fixed values of $N_{\beta}$ and $N_{\sigma}$, the function $\varepsilon(T)/T^{4}$ has two maxima, one maximum is at the lower temperatures and the other maximum is at the higher temperatures. See the left panels of Fig.~\ref{fig6}. This function vanishes as $T\to\infty$. Thus, the lattice physical energy density (\ref{72}) does not recover the Stefan-Boltzmann energy density $\varepsilon_{SB}$ at high temperatures. The second maximum of the function $\varepsilon(T)/T^{4}$ moves toward higher temperatures with $N_{\sigma}$ and the values of the function $\varepsilon(T)/T^{4}$ at the intermediate points between these two maxima tend to the values of energy in the continuum limit as $N_{\sigma}\to\infty$. The same behavior is clearly seen for the lattice physical thermodynamic potential $\omega_{ph}(T)/T^{4}$. However, the lattice physical pressure $p(T)/T^{4}$ has only one maximum at high temperatures and does not recover its Stefan-Boltzmann limit. The lattice trace anomaly $\Delta_{2}(T)$ has a maximum at low temperature and vanishes with increasing $T$. However, the lattice potential inhomogeneity $-\Delta_{1}(T)$ is an oscillating function of $T$ and it is not equal to zero for those values of $T$ for which the energy density and pressure are nonvanishing functions. Thus, as the lattice potential inhomogeneity $\Delta_{1}$ is not equal to zero, the lattice physical thermodynamic potential $(\Omega_{ph}=V\omega_{ph})$ is not a homogeneous function of the first order with respect to the extensive variable of state $V$ for all values of the temperature $T$ at fixed $N_{\beta}$, $N_{\sigma}$ and $V$. In this case the function $\omega_{ph}$ depends on $V$. Note that for $N_{\beta}=20$ and $N_{\sigma}<100$ at $V=3^{3}$ fm$^{3}$ and $T>200$ MeV all the lattice quantities differ from their corresponding continuum limit values.

The volume dependence of the physical energy density (\ref{72}), the physical pressure (\ref{73}), the density of the physical thermodynamic potential (\ref{71}), the trace anomaly (\ref{tt3}) and the potential inhomogeneity (\ref{pp3}) is presented in the right panels of Fig.~\ref{fig6}. The physical energy density $\varepsilon(V)$ on the finite lattice increases infinitely with the decrease in volume $V$. Such behaviour is also observed in the density of the lattice physical thermodynamic potential $\omega_{ph}(V)$. However, the lattice physical pressure $p(V)$ tends to zero as $V\to 0$. With growing volume $V$ the function $\varepsilon(V)$ first decreases, then increases and reaches its maximum and drops to zero in the limit $V\to\infty$. The maximum of the function $\varepsilon(V)$ moves toward larger volumes of $V$ with the increase of the lattice size $N_{\sigma}$. And the values of the function $\varepsilon(V)$ at the intermediate points between these two maxima tend to the values of energy in the continuum limit as $N_{\sigma}\to\infty$. The same behavior is also seen for the lattice physical thermodynamic potential $\omega_{ph}(V)$. The lattice physical pressure $p(V)$ increases with $V$, reaches its maximum and then drops to zero in the limit $V\to\infty$. The lattice trace anomaly $\Delta_{2}(V)$ and the lattice potential inhomogeneity $-\Delta_{1}(V)$ increase infinitely as $V\to 0$. The trace anomaly $\Delta_{2}$ and the potential inhomogeneity $\Delta_{1}$ on the finite lattice are not equal to zero for all values of $V$ for which the lattice physical energy density, the lattice physical thermodynamic potential and the lattice physical pressure are nonvanishing functions. Thus, as the lattice potential inhomogeneity $\Delta_{1}$ is not equal to zero, the lattice physical thermodynamic potential $\Omega_{ph}$ is not a homogeneous function of the first order. In the thermodynamic limit as $V\to\infty$ the lattice physical pressure (\ref{73}) and the lattice density of the physical thermodynamic potential (\ref{71}) vanish. Note that for $N_{\beta}=20$ and $N_{\sigma}<100$ at $T=100$ MeV and $V>9^{3}$ fm$^{3}$ all the lattice quantities differ from their corresponding continuum limit values.

\begin{figure}
\includegraphics[width=0.9\textwidth]{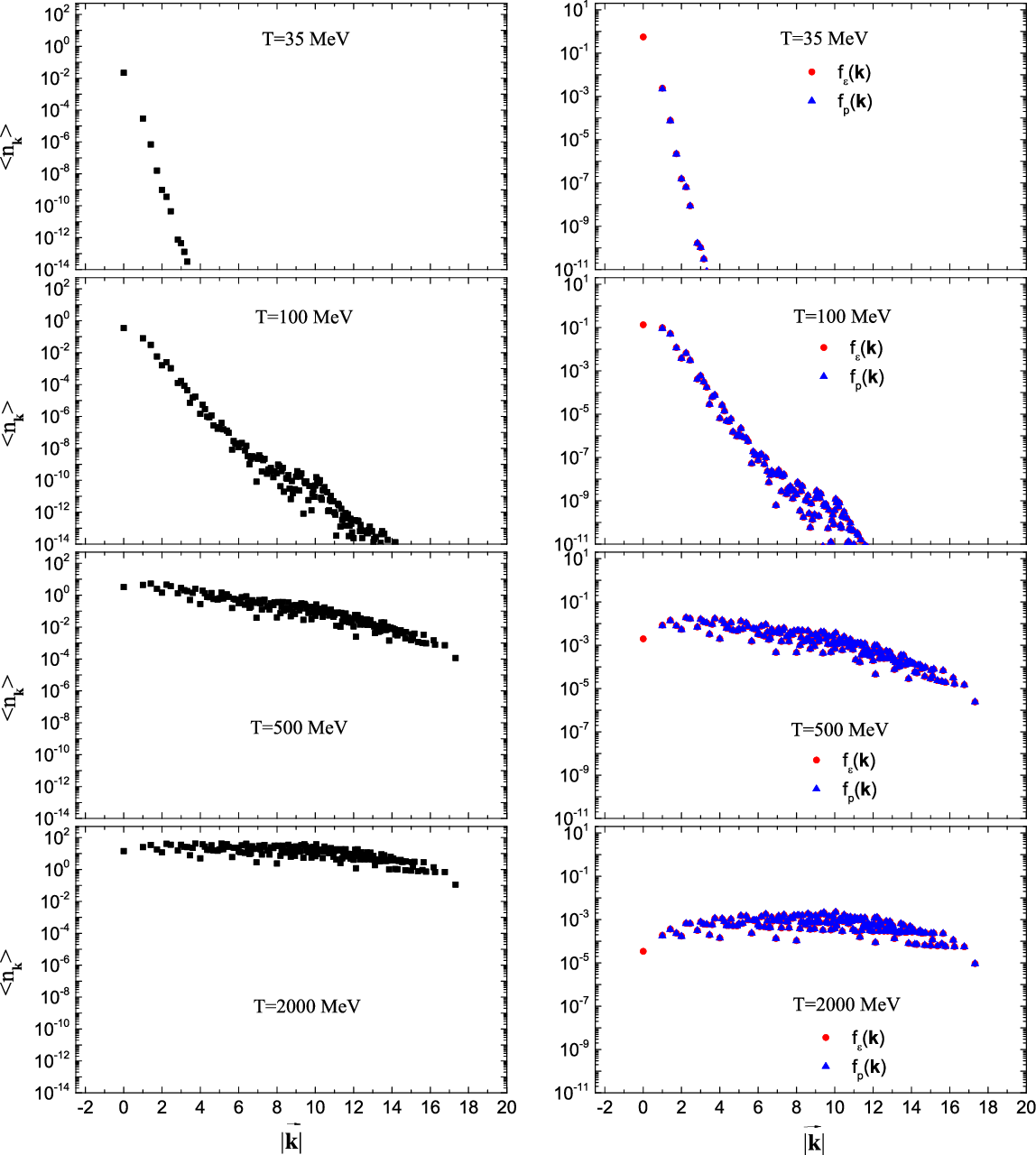}
\caption{(Color online) The mean occupation numbers $\langle n_{\vec{k}}\rangle$ (\ref{n4}), the distribution function $f_{\varepsilon}(\vec{k})$ (\ref{n10}) and the distribution function $f_{p}(\vec{k})$ (\ref{n11}) as functions of the length of the wavenumber vector $|\vec{k}|$ for the free neutral scalar field with the mass of $\pi^{0}$ pion on the lattice for $N_{\beta}=20$, $N_{\sigma}=20$ in the volume $V=3^{3}$ fm$^{3}$ for different values of temperature $T$.} \label{fig6a}
\end{figure}

\begin{figure}
\includegraphics[width=0.85\textwidth]{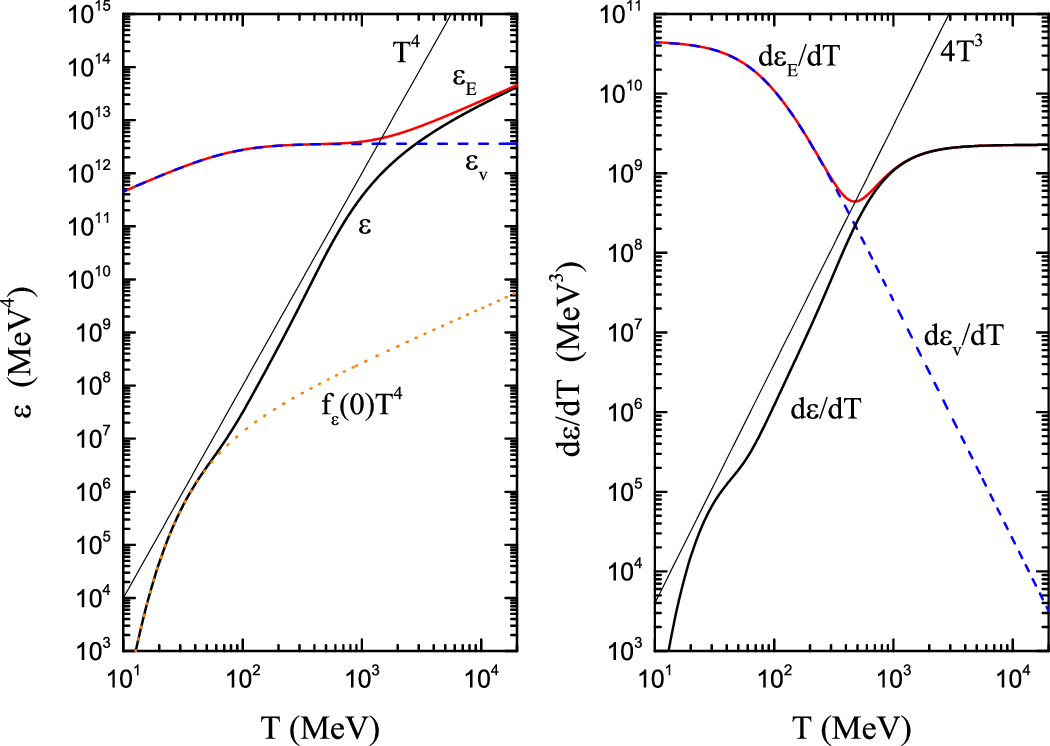}
\caption{(Color online) The energy density $\varepsilon_{E}$ (\ref{38}), the vacuum energy density $\varepsilon_{v}$ (\ref{69}), the physical energy density $\varepsilon=\varepsilon_{E}-\varepsilon_{v}$ (\ref{72}), their derivatives with respect to $T$ and the zero-mode term $f_{\varepsilon}(\vec{k}=0)T^{4}$ (\ref{n12}) as functions of the temperature $T$ for the free neutral scalar field with the mass of $\pi^{0}$ pion on the lattice for $N_{\beta}=20$, $N_{\sigma}=20$ in the volume $V=3^{3}$ fm$^{3}$.} \label{fig6b}
\end{figure}

Let us, now, explain why in Fig.~\ref{fig6} we have two maxima in $\varepsilon/T^{4}$ but only one in the pressure $3p/T^{4}$ at finite volume and why the pressure $3p/T^{4}$ vanishes for high temperature or for infinite volume on the fixed finite lattice. Let us rewrite Eqs.~(\ref{72}), (\ref{73}) and (\ref{tt3}) in the form
\begin{eqnarray}\label{n1}
  \varepsilon &=& \frac{1}{V} \sum\limits_{\vec{k}=k_{\sigma 1}}^{k_{\sigma 2}} \omega \ \langle n_{\vec{k}}\rangle, \\ \label{n2}
  p &=& \frac{1}{3V} \sum\limits_{\vec{k}=k_{\sigma 1}}^{k_{\sigma 2}} \left(\omega-\frac{m^{2}}{\omega}\right) \ \langle n_{\vec{k}}\rangle,\\ \label{n3}
  \Delta_{2} &=& \frac{1}{T^{4}V} \sum\limits_{\vec{k}=k_{\sigma 1}}^{k_{\sigma 2}} \frac{m^{2}}{\omega} \ \langle n_{\vec{k}}\rangle,
\end{eqnarray}
where
\begin{eqnarray}\label{n4}
  \langle n_{\vec{k}}\rangle &\equiv&  T \sum\limits_{k_{\beta}=k_{\beta 1}}^{k_{\beta 2}} \frac{\omega}{\omega_{\beta}^{2} + \omega^{2}} - \frac{1}{2} \left[1+\left(\frac{\omega}{2TN_{\beta}}\right)^{2}\right]^{-1/2}, \\ \label{n5}
  \omega_{\beta} &=& 2TN_{\beta} \sin \frac{\pi k_{\beta}}{N_{\beta}},\\ \label{n6}
  \omega &=& \sqrt{\sum\limits_{\alpha}\left( \frac{2N_{\sigma}}{V^{1/3}} \sin \frac{\pi k_{\alpha}}{N_{\sigma}} \right)^{2} + m^{2}}.
\end{eqnarray}
In the limit $N_{\beta}\to\infty,a_{\beta}\to 0$ at $\beta=const$ and $N_{\sigma}=const$ Eq.~(\ref{n4}) can be written as
\begin{equation}\label{n7}
  \langle n_{\vec{k}}\rangle^{*} = T \omega \sum\limits_{k_{\beta}=-\infty}^{\infty} \frac{1}{(2\pi Tk_{\beta})^{2} + \omega^{2}} - \frac{1}{2}.
\end{equation}
Using Eq.~(\ref{43}) we obtain
\begin{equation}\label{n8}
  \langle n_{\vec{k}}\rangle^{*} = \frac{1}{2} \coth \frac{\omega}{2T} - \frac{1}{2} = \frac{1}{e^{\omega/T}-1},
\end{equation}
where $\omega$ is calculated by Eq.~(\ref{n6}). In the continuum limit Eqs.~(\ref{n8}) and (\ref{n6}) take the form
\begin{equation}\label{n9}
  \langle n_{\vec{k}}\rangle^{c} = \frac{1}{e^{\omega/T}-1}, \qquad  \omega = \sqrt{\left(\frac{2\pi}{V^{1/3}}\right)^{2}\sum\limits_{\alpha} k_{\alpha}^{2} + m^{2}}.
\end{equation}
The quantity $\langle n_{\vec{k}}\rangle^{c}$ is the mean occupation numbers for the ideal boson gas in a finite volume in the continuum limit. Thus, the quantity $\langle n_{\vec{k}}\rangle$ may be considered like the mean occupation numbers for the neutral scalar field on the finite lattice. To understand the behaviour of the energy density and the pressure we should study all the terms of the sums (\ref{n1}) and (\ref{n2}). Thus, let us introduce the two distribution functions corresponding to the energy density $\varepsilon/T^{4}$ and the pressure $3p/T^{4}$:
\begin{equation}\label{n10}
  f_{\varepsilon}(\vec{k}) = \frac{\omega}{T^{4}V}\langle n_{\vec{k}}\rangle, \qquad \frac{\varepsilon}{T^{4}} = \sum\limits_{\vec{k}=k_{\sigma 1}}^{k_{\sigma 2}} f_{\varepsilon}(\vec{k})
\end{equation}
and
\begin{equation}\label{n11}
  f_{p}(\vec{k}) = \frac{1}{T^{4}V}\left(\omega-\frac{m^{2}}{\omega}\right) \ \langle n_{\vec{k}}\rangle, \qquad \frac{3p}{T^{4}} = \sum\limits_{\vec{k}=k_{\sigma 1}}^{k_{\sigma 2}} f_{p}(\vec{k}).
\end{equation}
Figure~\ref{fig6a} represents the behavior of the mean occupation numbers $\langle n_{\vec{k}}\rangle$ (\ref{n4}), the distribution function $f_{\varepsilon}(\vec{k})$ (\ref{n10}) and the distribution function $f_{p}(\vec{k})$ (\ref{n11}) as functions of the length of the wavenumber vector $|\vec{k}|$ for the free neutral scalar field with the mass of $\pi^{0}$ pion on the lattice for $N_{\beta}=20$, $N_{\sigma}=20$ in the volume $V=3^{3}$ fm$^{3}$ for different values of temperature $T$. Note that the calculations on Fig.~\ref{fig6a} correspond to the curves depicted by the number $1$ on the left upper panel of Fig.~\ref{fig6}. The first maximum in $\varepsilon/T^{4}$ corresponds to the temperature $T\approx 35$ MeV. At this value of temperature the distribution functions $\langle n_{\vec{k}}\rangle$ sharply decreases with $|\vec{k}|$ and the main contribution to the first maximum of $\varepsilon/T^{4}$ is provided by the zero-mode term $k_{x}=k_{y}=k_{z}=0$. The other terms $\vec{k}\neq 0$ are insignificant. It is not difficult to verify numerically that the rise and fall of the function $\varepsilon/T^{4}$ around its first maximum are very well described by the zero-mode term
\begin{equation}\label{n12}
  f_{\varepsilon}(\vec{k}=0)= \frac{m^{2}}{T^{3}V} \sum\limits_{k_{\beta}=k_{\beta 1}}^{k_{\beta 2}} \frac{1}{\omega_{\beta}^{2} + m^{2}} - \frac{m}{2T^{4}V} \left[1+\left(\frac{m}{2TN_{\beta}}\right)^{2}\right]^{-1/2}.
\end{equation}
See the left panel of Fig.~\ref{fig6b}. Contrariwise, the zero-mode term $k_{x}=k_{y}=k_{z}=0$ in the distribution $f_{p}(\vec{k})$ is equal to zero, $f_{p}(\vec{k}=0)=0$, because $\omega=m$ and $\omega-(m^{2}/\omega)=0$. Since the other terms $\vec{k}\neq 0$ are insignificant in the distribution $\langle n_{\vec{k}}\rangle$ we have $f_{p}(\vec{k})\sim 0$ and the pressure $3p/T^{4}\sim 0$. See the upper panels of Fig.~\ref{fig6a} and the left upper panel of Fig.~\ref{fig6}. Thus, the vanishing zero-mode term in the pressure for the massive scalar field leads to the vanishing pressure in the temperature range around the first maximum of $\varepsilon/T^{4}$. This leads to the nonvanishing trace anomaly $\Delta_{2}\approx f_{\varepsilon}(\vec{k}=0)$ and the nonvanishing potential inhomogeneity, which is not equal to zero due to the nonvanishing density of the physical thermodynamic potential (\ref{71}). See the left lower panel of Fig.~\ref{fig6}. Thus, in the temperature range around the first maximum of $\varepsilon/T^{4}$, the trace anomaly $\Delta_{2}\approx \varepsilon/T^{4}$ and the nonvanishing potential inhomogeneity for the massive scalar field on the finite lattice are related to the overwhelming contribution of the zero-mode term $k_{x}=k_{y}=k_{z}=0$ to these quantities. The distribution function $f_{p}(\vec{k})$ practically coincides with $f_{\varepsilon}(\vec{k})$ for all $\vec{k}\neq 0$ at any value of temperature $T$. The contribution of the zero-mode term in $f_{\varepsilon}(\vec{k})$ and $\langle n_{\vec{k}}\rangle$ decreases with $T$. Therefore, at high temperatures the pressure $3p/T^{4}$ becomes identical with the energy density $\varepsilon/T^{4}$. The appearance of the second maximum in $\varepsilon/T^{4}$ and the maximum in $3p/T^{4}$ may be explained by the finiteness of the $3$-dimensional momentum space and the limited values of $k_{\beta}$ for the fixed finite lattice $(N_{\beta},N_{\sigma})$. This maximum shifts toward higher temperatures with $N_{\sigma}$ and completely disappears in the continuum limit. See the left upper panel of Fig.~\ref{fig6}. In the continuum limit the $3$-dimensional momentum space becomes infinitely large and the range of $k_{\beta}$ also grows to infinity. In detail the second maximum in $\varepsilon/T^{4}$ is determined by the inflection point in the energy density $\varepsilon_{E}$ (\ref{38}), when the derivative $\partial \varepsilon_{E}/\partial T$ has the minimum, and by the achievement of the vacuum energy density its maximal constant value. See Fig.~\ref{fig6b}. The inflection point in the energy density $\varepsilon_{E}$ appears due to the increase of $\omega_{\beta}$, which depends on $N_{\beta}$ and $T$, with $T$ at constant value of $\omega$, which depends solely on $N_{\sigma}$ and $V$, and does not depend on the temperature $T$.  With the growth of the temperature $T$ the behaviour of the physical energy density $\varepsilon$ is determined by the energy density $\varepsilon_{E}$ since the vacuum energy density $\varepsilon_{v}$ achieves its maximal constant value. Since the $3$-momentum is confined $k_{\sigma 1}\leq \vec{k} \leq k_{\sigma 2}$ at the fixed finite value of $N_{\sigma}$ the distribution $\langle n_{\vec{k}}\rangle$ becomes more uniform and grows with $T$. See the lowest left two panels of Fig.~\ref{fig6a}. This leads to the increase of $\varepsilon$ and $p$ with $T$. The quantities $\langle n_{\vec{k}}\rangle$, $\varepsilon$ and $p$ grow linearly with $T$ as $T\to\infty$ because in the sum with regard to $k_{\beta}$ in Eq.~(\ref{n4}) only the term $k_{\beta}=0$ survives, i.e. $\langle n_{\vec{k}}\rangle=T/\omega$, $\varepsilon=TN_{\sigma}^{3}/V$ and $p=T(N_{\sigma}^{3}-\sum_{\vec{k}=k_{\sigma 1}}^{k_{\sigma 2}}m^{2}/\omega^{2})/3V$. Therefore, the energy density $\varepsilon/T^{4}\to 0$ and the pressure $3p/T^{4}\to 0$ as $T\to\infty$. In the limit $V\to\infty$ at $T=const$ we have $\omega=m$ and $\langle n_{\vec{k}}\rangle=\langle n_{\vec{k}=0}\rangle$ for all $\vec{k}$. This leads to $\varepsilon=m\langle n_{\vec{k}=0}\rangle N_{\sigma}^{3}/V\to 0$ and $p=(m-m)\langle n_{\vec{k}=0}\rangle N_{\sigma}^{3}/3V\to 0$ as $V\to\infty$.

\begin{figure*}
\includegraphics[width=0.93\textwidth]{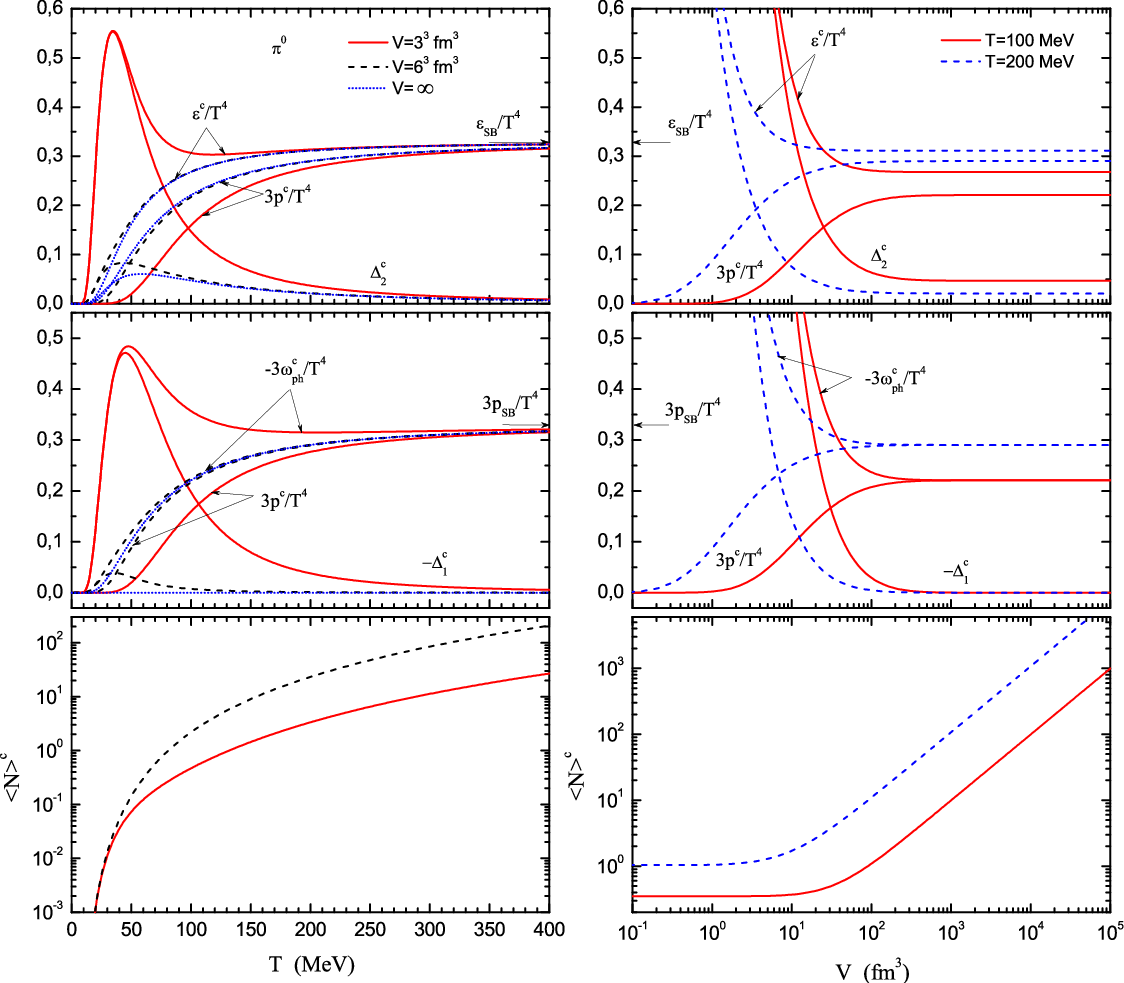}
\caption{(Color online) Continuum limit for the massive scalar field. The physical energy density $\varepsilon^{c}$ (\ref{60}), the physical pressure $p^{c}$ (\ref{64}), the density of the physical thermodynamic potential $\omega_{ph}^{c}$ (\ref{67}), the trace anomaly $\Delta_{2}^{c}$ (\ref{tt1}), the potential inhomogeneity $\Delta_{1}^{c}$ (\ref{pp1}) and the mean number of particles $\langle N\rangle^{c}$ as functions of volume $V$ and temperature $T$ for the free neutral scalar field with the mass of $\pi^{0}$ pion in the continuum limit. Left panels: the curves were calculated at the volume $V=3^{3}$, $6^{3}$ fm$^{3}$ and in the thermodynamic limit $(V\to\infty)$. Right panels: the curves correspond to the temperature $T=100$ and $200$ MeV.} \label{fig7}
\end{figure*}

Figure~\ref{fig7} represents the behavior of the physical energy density (\ref{60}), the physical pressure (\ref{64}), the density of the physical thermodynamic potential (\ref{67}), the trace anomaly (\ref{tt1}), the potential inhomogeneity (\ref{pp1}) and the mean number of particles $\langle N\rangle^{c}=\sum_{\vec{p}} 1/(e^{\beta \omega}-1)$ as functions of volume $V$ and temperature $T$ for the free neutral scalar field in the continuum limit with the mass of $\pi^{0}$ pion. For $V=3^{3}$ fm$^{3}$, the function $\varepsilon^{c}(T)/T^{4}$ has a maximum at $T\sim 30-40$ MeV. With the growth of the temperature $T$ this function decreases and attains a constant value as $T\to\infty$. For $V=6^{3}$ fm$^{3}$ and in the thermodynamic limit, the function $\varepsilon^{c}(T)/T^{4}$ is a monotonically increasing function with $T$ and tends to the same constant value as $T\to\infty$. The same behavior is seen for the physical thermodynamic potential $-\omega_{ph}^{c}(T)/T^{4}$; however, the maximum of this function is located at $T\sim 50$ MeV. For $V=3^{3}$ fm$^{3}$, $V=6^{3}$ fm$^{3}$ and in the thermodynamic limit, the physical pressure $p^{c}(T)/T^{4}$ increases monotonically with $T$ and tends to a constant value as $T\to\infty$. The continuum physical energy density $\varepsilon^{c}$, the continuum physical thermodynamic potential $\omega_{ph}^{c}$ and the continuum physical pressure $p^{c}$ tend to their thermodynamic limit values with $T$ and resemble the Stefan-Boltzmann limit as $T\to\infty$. For all values of $V$, the continuum trace anomaly $\Delta_{2}^{c}(T)$ has a maximum. With the growth of the temperature $T$ this function decreases and tends to zero as $T\to\infty$. The appearance of the maximum in the trace anomaly in the continuum limit at low temperatures $T$ is independent of the presence of the maximum in the energy density. For both $V=3^{3}$ fm$^{3}$ and $V=6^{3}$ fm$^{3}$, the potential inhomogeneity $-\Delta_{1}^{c}(T)$ in the continuum limit has a maximum. However $\Delta_{1}^{c}$ is equivalent to zero for $V\to\infty$. The potential inhomogeneity $-\Delta_{1}^{c}(T)$ for the free massive neutral scalar field in the continuum limit decreases with $T$ and tends to zero as $T\to\infty$. The nonzero values of the potential inhomogeneity $\Delta_{1}^{c}$ at finite values of $T$ indicate that for fixed values of $V$ the continuum physical thermodynamic potential $(\Omega_{ph}^{c}=V\omega_{ph}^{c})$ is not a homogeneous function of the first order with respect to the extensive variable of state $V$. The function $\omega_{ph}^{c}$ depends on $V$.

The volume dependence of the physical energy density (\ref{60}), the physical pressure (\ref{64}), the density of the physical thermodynamic potential (\ref{67}), the trace anomaly (\ref{tt1}), the potential inhomogeneity (\ref{pp1}) and the mean number of particles $\langle N\rangle^{c}$ in the continuum limit is presented in the right panels of Fig.~\ref{fig7}. The physical energy density $\varepsilon^{c}$ in the continuum limit increases infinitely with the decrease in volume $V$. Such behaviour is also seen in the density of the continuum physical thermodynamic potential $\omega_{ph}^{c}$. However, the physical pressure $p^{c}$ tends to zero as $V\to 0$. With the growth of the volume towards $V>10^{3}$ fm$^{3}$ the physical energy density $\varepsilon^{c}$, the physical thermodynamic potential $\omega_{ph}^{c}$ and the physical pressure $p^{c}$ in the continuum limit reach their values in the thermodynamic limit, which are different from their values in the Stefan-Boltzmann limit. The trace anomaly $\Delta_{2}^{c}$ and the potential inhomogeneity $-\Delta_{1}^{c}$ increase infinitely as $V\to 0$. The trace anomaly $\Delta_{2}^{c}$ in the continuum limit is not equal to zero for all values of $V$ and fixed temperature $T$. The potential inhomogeneity $-\Delta_{1}^{c}$ for the free massive neutral scalar field in the continuum limit decreases with $V$ and vanishes in the thermodynamic limit as $V\to\infty$. Thus, in the thermodynamic limit the continuum physical thermodynamic potential $\Omega_{ph}^{c}$ is a homogeneous function of the first order with respect to the extensive variable of state $V$.

\begin{figure*}
\includegraphics[width=0.93\textwidth]{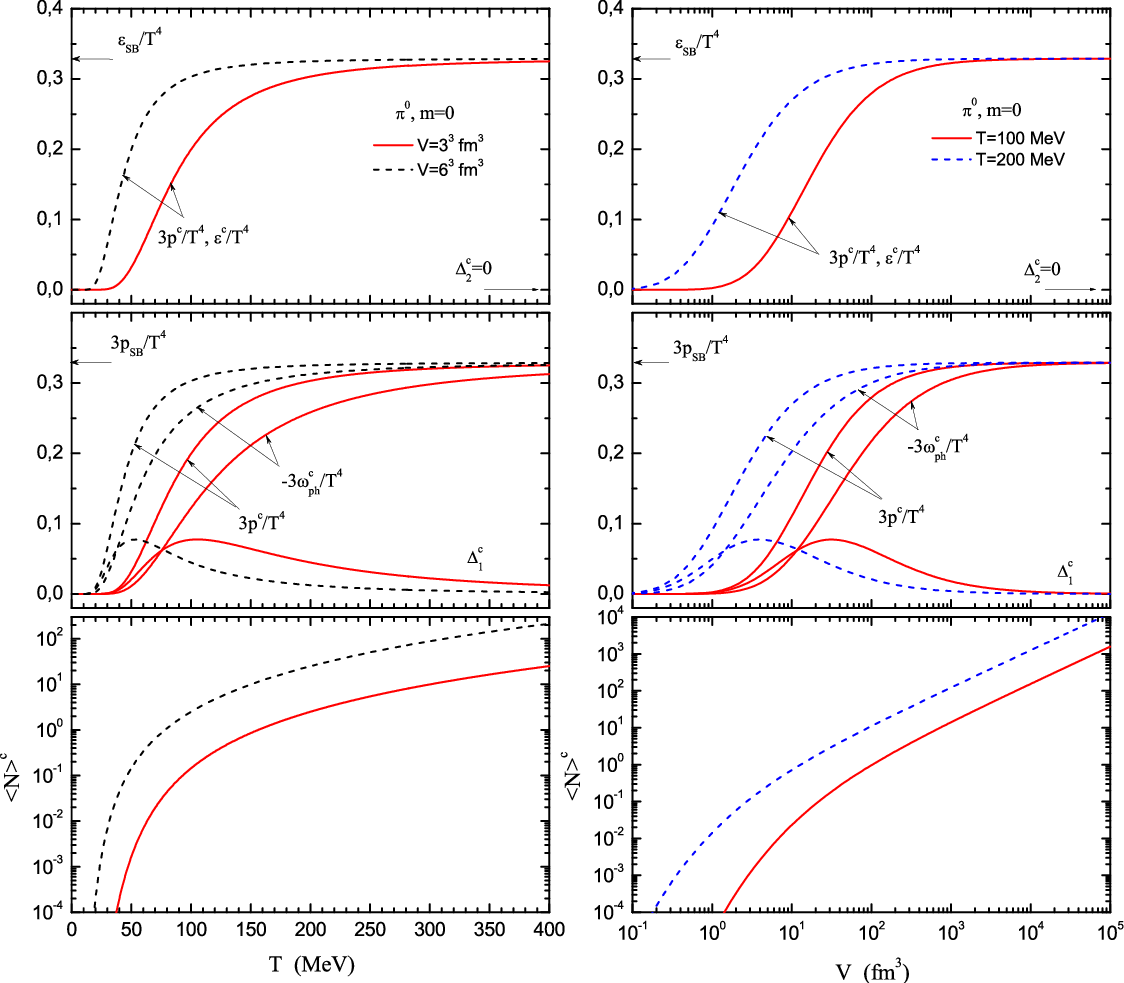}
\caption{(Color online) Continuum limit for the massless scalar field without zero-mode term. The physical energy density $\varepsilon^{c}$ (\ref{60}), the physical pressure $p^{c}$ (\ref{64}), the density of the physical thermodynamic potential $\omega_{ph}^{c}$ (\ref{67}), the trace anomaly $\Delta_{2}^{c}$ (\ref{tt1}), the potential inhomogeneity $\Delta_{1}^{c}$ (\ref{pp1}) and the mean number of particles $\langle N\rangle^{c}$ as functions of volume $V$ and temperature $T$ for the free neutral scalar field with the mass $m=0$ in the continuum limit without zero-mode term. Left panels: the curves were calculated at the volume $V=3^{3}$ and $6^{3}$ fm$^{3}$. Right panels: the curves correspond to the temperature $T=100$ and $200$ MeV.} \label{fig8}
\end{figure*}

Figure~\ref{fig8} represents the behavior of the physical energy density (\ref{60}), the physical pressure (\ref{64}), the density of the physical thermodynamic potential (\ref{67}), the trace anomaly (\ref{tt1}), the potential inhomogeneity (\ref{pp1}) and the mean number of particles $\langle N\rangle^{c}$ as functions of volume $V$ and temperature $T$ for the free massless neutral scalar field in the continuum limit without zero-mode term. For $V=3^{3}$ and $V=6^{3}$ fm$^{3}$, the physical energy density $\varepsilon^{c}(T)/T^{4}$ coincides with the physical pressure $p^{c}(T)/T^{4}$ and they both are monotonically increasing functions with $T$. The same behavior is seen for the physical thermodynamic potential $-\omega_{ph}^{c}(T)/T^{4}$; however, the physical thermodynamic potential $-\omega_{ph}^{c}(T)/T^{4}$ does not coincide with the physical pressure $p^{c}(T)/T^{4}$. The physical energy density $\varepsilon^{c}$, the physical thermodynamic potential $\omega_{ph}^{c}$ and the physical pressure $p^{c}$ in the continuum limit tend to the Stefan-Boltzmann energy density $\varepsilon_{SB}$, the Stefan-Boltzmann density of the thermodynamic potential $\omega_{SB}$ and the Stefan-Boltzmann pressure $p_{SB}$, respectively, as $T\to\infty$. For all values of $V$, the trace anomaly $\Delta_{2}^{c}(T)$ for the massless field in the continuum limit is equal to zero. For both $V=3^{3}$ fm$^{3}$ and $V=6^{3}$ fm$^{3}$, the potential inhomogeneity $-\Delta_{1}^{c}(T)$ for the free massless neutral scalar field in the continuum limit has a maximum. It decreases with $T$ and tends to zero as $T\to\infty$. The nonzero values of the potential inhomogeneity $\Delta_{1}^{c}$ at finite values of $T$ indicate that for fixed values of $V$ the continuum physical thermodynamic potential $\Omega_{ph}^{c}$ is not a homogeneous function of the first order.

The volume dependence of the physical energy density (\ref{60}), the physical pressure (\ref{64}), the density of the physical thermodynamic potential (\ref{67}), the trace anomaly (\ref{tt1}), the potential inhomogeneity (\ref{pp1}) and the mean number of particles $\langle N\rangle^{c}$ for the free massless neutral scalar field in the continuum limit is presented in the right panels of Fig.~\ref{fig8}. The physical energy density $\varepsilon^{c}$ coincides with the physical pressure $p^{c}$ and they both tend to zero as $V\to 0$. The physical thermodynamic potential $\omega_{ph}^{c}$ in the continuum limit also tends to zero as $V\to 0$; however, it does not coincide with the physical pressure $p^{c}$ at finite $V$. With the growth of the volume towards $V>10^{4}$ fm$^{3}$ the physical energy density $\varepsilon^{c}$, the physical thermodynamic potential $\omega_{ph}^{c}$ and the physical pressure $p^{c}$ in the continuum limit reach their values in the thermodynamic limit, which correspond also to the values of the Stefan-Boltzmann limit. The trace anomaly $\Delta_{2}^{c}$ in the continuum limit is equal to zero for all values of $V$ and fixed temperature $T$. The potential inhomogeneity $-\Delta_{1}^{c}$ for the free massless neutral scalar field in the continuum limit decreases with $V$ and vanishes in the thermodynamic limit as $V\to\infty$. Thus, the zero values of the potential inhomogeneity $\Delta_{1}^{c}$ in the thermodynamic limit indicate that for fixed values of $T$ the continuum physical thermodynamic potential $\Omega_{ph}^{c}$ is a homogeneous function of the first order.

In Fig.~\ref{fig8}, it was shown numerically that the physical pressure (\ref{64}) for the free massless neutral scalar field in the continuum limit without zero-mode term vanishes in the limit as $T\to 0$ or $V\to 0$. Let us prove this statement analytically. Let us rewrite the physical pressure (\ref{64}) for $m=0$ in a explicit form
\begin{eqnarray}\label{m2}
  p^{c} &=& \frac{1}{3} \sum\limits_{k_{x}=-\infty}^{\infty} \sum\limits_{k_{y}=-\infty}^{\infty} \sum\limits_{k_{z}=-\infty}^{\infty} f(k_{x},k_{y},k_{z}), \qquad f(k_{x},k_{y},k_{z}) = \frac{1}{V} \frac{\Delta p \sqrt{k_{x}^{2}+k_{y}^{2}+k_{z}^{2}}}{e^{\beta \Delta p \sqrt{k_{x}^{2}+k_{y}^{2}+k_{z}^{2}}}-1},
\end{eqnarray}
where $\Delta p=2\pi/V^{1/3}$. In this paper we consider only the case without zero-mode term for the free massless neutral scalar field. Thus, the zero-mode term $(k_{x}=k_{y}=k_{z}=0)$ in Eq.~(\ref{m2}) was suppressed by definition. Since $\vec{k}^{2}=k_{x}^{2}+k_{y}^{2}+k_{z}^{2}\neq 0$ for all terms of the sum (\ref{m2}), we have $f(\vec{k})=(\Delta p)^{4} |\vec{k}| e^{-\Delta p |\vec{k}|/T}/(2\pi)^{3}\to 0$ as $T\to 0$ and $V=const$ $(\Delta p = const)$ or $V\to 0$ $(\Delta p \to \infty)$ and $T=const$. Thus the physical pressure (\ref{m2}) and the ratio $3p^{c}/T^{4}$ vanish in the limit as $T\to 0$ or $V\to 0$.

In the continuum limit the nonvanishing potential inhomogeneity $(\Delta_{1}^{c}\ne 0)$ of the system at low temperatures $T$ and small values of volume $V$ may be explained by the fact that in the volume $V$ the system contains less than one particle on average, $\langle N\rangle^{c}\leq 1$. See the lower panels of Figs.~\ref{fig7} and \ref{fig8}. In the grand canonical ensemble the term with the zero number of particles also contributes to the partition function. Note that for the free real scalar field there is no conserved current. However, the number of neutral particles can be accounted by the eigenvalues of the number operator
\begin{equation}\label{m1}
 N=\int d^{3}p a^{+}(p)a(p).
\end{equation}
The definition of this operator for the free real scalar field can be found, for example, in~\cite{Kaku}. The mean number of particles in the continuum limit $\langle N\rangle^{c}$ corresponds, namely, to the statistical average of this number operator $N$ and has the meaning of the average number of neutral bosons in the volume $V$ at temperature $T$~\cite{Kapusta89}. With the growth of the volume $V$ and temperature $T$ the mean number of particles in the continuum limit $\langle N\rangle^{c}$ increases and the thermodynamic potential becomes a homogeneous function, i.e., $\Delta_{1}^{c}\to 0$.

\section{Discussion and conclusions}\label{sec6}
In the present paper, the partition function for the free neutral scalar field in one spatial dimension with the periodic and antiperiodic boundary conditions along the spatial axis was exactly derived in a general form in the framework of the path integral method in both configuration space and momentum space. The path integral method used in this paper provides the same results for the partition function of the scalar field as the standard method of path integrals given, for example, in Ref.~\cite{Engels}. The symmetric square matrices $A$ and $B$ of the bilinear forms on the vector space of fields in both the configuration and the momentum space were found explicitly. The transformation to the momentum space in one spatial dimension was implemented by the lattice Fourier transform. To solve exactly the partition function in the configuration space the recurrence equations were obtained. The exact analytical results for the partition function in one spatial dimension in the momentum space were also obtained. It was numerically proved that the partition functions in the configuration space and the momentum space are equivalent. The partition function was generalized to the three-dimensional spatial momentum space. The main thermodynamic quantities of the grand canonical ensemble on the finite lattice in the three-dimensional spatial momentum space were derived from the partition function with the periodic and antiperiodic boundary conditions along the spatial axes. In a particular case of the periodic spatial boundary conditions we recovered the results firstly obtained in~\cite{Engels} by another method.

We have found the exact analytical expressions for the thermodynamic quantities of the free neutral scalar field in the continuum limit and in the limit $N_{\beta}\to\infty$, $\beta=const$ and $N_{\sigma}=const$. In both limits the thermodynamic quantities of the free neutral scalar field are split into two parts, the sum of the vacuum and physical terms. In the continuum limit all the vacuum terms are numerically divergent. However, in the limit $N_{\beta}\to\infty$, $\beta=const$ and $N_{\sigma}=const$ the vacuum terms are finite numbers. In the continuum limit, the thermodynamic quantities of the free neutral scalar field obtained by the method of path integrals used in this paper exactly coincide with their corresponding quantities obtained by the method of canonical quantization. In particular, in the thermodynamic limit the continuum thermodynamic quantities for the massless neutral scalar field recover their Stefan-Boltzmann limit values.

The vacuum and physical terms of the thermodynamic quantities on a finite lattice were defined as in Ref.~\cite{Engels}. In the continuum limit, these lattice quantities exactly coincide with their corresponding continuum quantities obtained by the method of path integrals considered in this paper. The same behaviour is clearly seen for them in the limit $N_{\beta}\to\infty$, $\beta=const$ and $N_{\sigma}=const$.

The principle of additivity for any statistical system in the grand canonical ensemble has been considered. We demonstrated in a general form that if the thermodynamic potential of the grand canonical ensemble is not a homogeneous function of the first order with respect to the extensive variable of state $V$, then the thermodynamic potential is nonadditive, the potential inhomogeneity is not equal to zero and, hence, the zeroth law of thermodynamics in the grand canonical ensemble is not satisfied.

The thermodynamic properties and the finite volume corrections to the thermodynamic quantities of the free real scalar field both on the finite lattice and in the continuum limit have been studied in the range of temperature and volume typical of the ultrarelativistic proton-proton and heavy-ion collisions. We showed that on the finite lattice the energy density $\varepsilon_{E}$ and the physical energy density $\varepsilon$ recover their limiting values $\varepsilon_{E}^{*}$ and $\varepsilon^{*}$, $\varepsilon^{c}$, respectively, with the increase of the sizes of the lattice at fixed finite values of temperature and volume. From the exact calculations of the lattice physical energy density and its continuum limit value we estimated the number of sites of the lattice which approximate the continuum limit value of the physical energy density with certain accuracy in the dependence of the volume of the system at fixed finite temperature $T$.

We have found that on the finite lattice the exact lattice results for the free massive neutral scalar field are in good agreement with the continuum limit only in the region of small values of temperature and volume, where the continuum physical thermodynamic potential of the grand canonical ensemble is not a homogeneous function of the first order with respect to the extensive variable of state $V$. At higher values of temperature and volume, where the continuum physical thermodynamic potential is a homogeneous function of the first order with respect to the extensive variable of state $V$, the lattice physical quantities deviate essentially from their continuum limit and have nonphysical behavior. Furthermore, the lattice physical thermodynamic potential on a finite lattice is not a homogeneous function of the first order with respect to the extensive variable of state $V$ in the whole range of the variables of state $T$ and $V$ because the density of the lattice physical thermodynamic potential is a function of volume.

We revealed that at low temperatures and small values of volume the continuum physical quantities for the massive scalar field deviate essentially from their thermodynamic limit values. With the growth of the temperature at fixed small values of volume the continuum physical quantities tend to their thermodynamic limit. Such behaviour in the continuum physical quantities is also seen with the growth of the volume at fixed temperature. At the same time the higher the temperature, the better the thermodynamic limit values of these quantities  approach the Stefan-Boltzmann limit. For the massive scalar field the continuum trace anomaly is not equal to zero at low temperatures and all values of volume. With the growth of the temperature at fixed values of volume, the continuum trace anomaly vanishes. The continuum potential inhomogeneity is not equal to zero at small values of volume and temperature. However, in the thermodynamic limit or/and at high temperatures it vanishes and the continuum physical thermodynamic potential becomes a homogeneous function of the first order with respect to the extensive variable of state $V$.

We found that at low temperatures and small values of volume the continuum physical quantities for the massless scalar field without zero-mode term deviate essentially from their thermodynamic limit which in the case of the massless field is the Stefan-Boltzmann limit. With a growing temperature at fixed small values of volume and with a growing volume at fixed temperature, the continuum physical quantities tend to their thermodynamic limit values. At the same time the higher the temperature and volume, the faster they tend to the thermodynamic limit with the increasing volume and temperature. The continuum trace anomaly for the massless scalar field without zero-mode term is equal to zero for all values of volume and temperature. The continuum potential inhomogeneity is not equal to zero at small values of volume and temperature. However, in the thermodynamic limit or/and at higher temperatures it vanishes. In this case the continuum physical thermodynamic potential becomes a homogeneous function of the first order with respect to the extensive variable of state $V$. It should be stressed that at low temperatures and small values of volume the continuum physical thermodynamic potentials for both the massive and the massless scalar field are not a homogeneous function of the first order because in this case the continuum mean number of particles in the system is less than one particle on average. In general, the continuum quantities of the ideal gas in a finite volume deviate essentially from their thermodynamic limit values due to the quantum effects related to the discrete energy spectra of particles.

{\bf Acknowledgments:} This work was supported in part by the joint research project and grant of JINR and IFIN-HH (protocols N~4342 and N~4543). I am indebted to E.-M.~Ilgenfritz and O.V.~Teryaev for valuable remarks and fruitful discussions.



\begin{thebibliography} {xxxxx}

\bibitem{Abelev13} B.~Abelev {\it et al}. (ALICE Collaboration), Phys. Rev. Lett. {\bf 110}, 082302 (2013).

\bibitem{Khachatryan10_1} V.~Khachatryan {\it et al}. (CMS Collaboration), Phys. Rev. Lett. {\bf 105}, 022002 (2010).

\bibitem{Chatrchyan13} S.~Chatrchyan {\it et al}. (CMS Collaboration), Phys. Rev. D {\bf 88}, 052001 (2013).

\bibitem{Kharlov13} Y.~Kharlov (ALICE Collaboration), Nucl. Phys. A {\bf 910-911}, 335 (2013).


\bibitem{Arsene05} I.~Arsene {\it et al}. (BRAHMS Collaboration), Phys. Rev. C {\bf 72}, 014908 (2005).

\bibitem{Adams06} J.~Adams {\it et al}. (STAR Collaboration), Phys. Lett. B {\bf 637}, 161 (2006).

\bibitem{Adare11} A.~Adare {\it et al}. (PHENIX Collaboration), Phys. Rev. D {\bf 83}, 052004 (2011).


\bibitem{Yagi} K.~Yagi, T.~Hatsuda, and Y.~Miake, {\it Quark-gluon plasma. From big bang to little bang} (Cambridge University Press, London, 2005).


\bibitem{Cleymans86} J.~Cleymans, R.V.~Gavai, and E.~Suhonen, Phys. Rep. {\bf 130}, 217 (1986).

\bibitem{BraunMunzinger09} P.~Braun-Munzinger and J.~Wambach, Rev. Mod. Phys. {\bf 81}, 1031 (2009).


\bibitem{Stocker86} H.~St\"{o}cker and W.~Greiner, Phys. Rep. {\bf 137}, 277 (1986).

\bibitem{Andronic12} A.~Andronic, P.~Braun-Munzinger, J.~Stachel, and M.~Winn, Phys. Lett. B {\bf 718}, 80 (2012).

\bibitem{Kapoyannis07} A.S.~Kapoyannis, Eur. Phys. J. C {\bf 51}, 135 (2007).

\bibitem{Lee93} K.S.~Lee and U.~Heinz, Phys. Rev. D {\bf 47}, 2068 (1993).

\bibitem{Kahara08} T.~K\"{a}h\"{a}r\"{a} and K.~Tuominen, Phys. Rev. D {\bf 78}, 034015 (2008).

\bibitem{Shao11} G.Y.~Shao, M.~Di~Toro, V.~Greco, M.~Colonna, S.~Plumari, B.~Liu, and Y. X.~Liu, Phys. Rev. D {\bf 84}, 034028 (2011).

\bibitem{Tiwari12} S.K.~Tiwari, P.K.~Srivastava, and C.P.~Singh, Phys. Rev. C {\bf 85}, 014908 (2012).


\bibitem{Braun13} J.~Braun, B.~Klein, and B.-J.~Schaefer, Phys. Lett. B {\bf 713}, 216 (2012).


\bibitem{Greiner96} W.~Greiner and J.~Reinhardt, {\it Field quantization} (Springer, Berlin Heidelberg, 1996).

\bibitem{Kapusta89} J.I.~Kapusta, {\it Finite-temperature field theory} (Cambridge University Press, London, 1989).


\bibitem{Greiner07} W.~Greiner, S.~Schramm, and E.~Stein, {\it Quantum Chromodynamics} (Springer, Berlin Heidelberg, 2007).

\bibitem{Strodthoff12} N.~Strodthoff, B.-J. Schaefer, and L. von Smekal, Phys. Rev. D {\bf 85}, 074007 (2012).


\bibitem{Muta98} T.~Muta, {\it Foundations of Quantum Chromodynamics: An Introduction to Perturbative Methods in Gauge Theories} (World Scientific, Singapore, 1998).

\bibitem{Gattringer10} C.~Gattringer and C.B.~Lang, {\it Quantum Chromodynamics on the Lattice: An Introductory Presentation}, Lect. Notes Phys. {\bf 788} (Springer, Berlin Heidelberg, 2010).


\bibitem{Prigogine} I.~Prigogine and D.~Kondepudi, {\it Modern Thermodynamics: From Heat Engines to Dissipative Structures} (John Wiley \& Sons, Chichester, 1998).

\bibitem{Parvan1} A.S.~Parvan, Nucl. Phys. A {\bf 887}, 1 (2012).

\bibitem{Parvan2} A.S.~Parvan, Phys. Lett. A {\bf 350}, 331 (2006).

\bibitem{Parvan3} A.S.~Parvan, Phys. Lett. A {\bf 360}, 26 (2006).

\bibitem{Parvan2015} A.S.~Parvan, Eur. Phys. J. A \textbf{51}, 108 (2015)

\bibitem{Parvan2015a} A.S.~Parvan, Foundation of equilibrium statistical mechanics based on generalized entropy, in {\it Recent Advances in Thermo and Fluid Dynamics}, ed. by Mofid Gorji-Bandpy (InTech, Rijeka, 2015), p.303


\bibitem{Boyd} G.~Boyd, J.~Engels, F.~Karsch, E.~Laermann, C.~Legeland, M.~L\"{u}tgemeier, B.~Petersson, Phys. Rev. Lett. {\bf 75}, 4169 (1995).

\bibitem{Borsanyi14} S.~Bors\'{a}nyi, Z.~Fodor, C.~Hoelbling, S.D.~Katz, S.~Kreig, K.K.~Szab\'{o}, Phys. Lett. B {\bf 730}, 99 (2014).

\bibitem{Borsanyi10} S.~Bors\'{a}nyi, G.~Endr\"{o}di, Z.~Fodor, A.~Jakov\'{a}c, S.D.~Katz, S.~Kreig, C.~Ratti, and K.K.~Szab\'{o}, JHEP {\bf 11}, 077 (2010).

\bibitem{Borsanyi13} S.~Bors\'{a}nyi, Z.~Fodor, C.~Hoelbling, S.D.~Katz, S.~Kreig, K.K.~Szab\'{o}, PoS LATTICE 2013, 155 (2013).

\bibitem{Aoki06} Y.~Aoki, Z.~Fodor, S.D.~Katz, and K.K.~Szab\'{o}, Phys. Lett. B {\bf 643}, 46 (2006).


\bibitem{Redlich} K.~Redlich, K.~Zalewski, arXiv:1611.03746 [nucl-th].


\bibitem{Engels} J.~Engels, F.~Karsch, and H.~Satz, Nucl. Phys. B {\bf 205}, 239 (1982).

\bibitem{Watkins} D.S.~Watkins, {\it Fundamentals of Matrix Computations} (John Wiley \& Sons, New York, 1991).

\bibitem{Gradshteyn} I.S.~Gradshteyn and I.M.~Ryzhik, {\it Table of Integrals, Series, and Products} (Academic Press, New York, 1994).


\bibitem{Bellac} M.~Le~Bellac, {\it Thermal Field Theory} (Cambridge University Press, London, 1996).


\bibitem{Abelev14} B.~Abelev {\it et al}. (ALICE Collaboration), Phys. Rev. C {\bf 89}, 024911 (2014).

\bibitem{Aamodt11} K.~Aamodt {\it et al}. (ALICE Collaboration), Phys. Rev. D {\bf 84}, 112004 (2011).

\bibitem{Khachatryan10_2} V.~Khachatryan {\it et al}. (CMS Collaboration), Phys. Rev. Lett. {\bf 105}, 032001 (2010).

\bibitem{Mercado11} J.~Mercado (ALICE Collaboration), J. Phys. G: Nucl. Part. Phys. {\bf 38}, 124056 (2011).

\bibitem{Mizoguchi10} T.~Mizoguchi and M.~Biyajima, Eur. Phys. J. C {\bf 70}, 1061 (2010).

\bibitem{Csorgo06} T.~Cs\"{o}rg\H{o}, J. Phys.: Conf. Ser. {\bf 50}, 259 (2006).


\bibitem{Adare10} A.~Adare {\it et al}. (PHENIX Collaboration), Phys. Rev. Lett. {\bf 104}, 132301 (2010).

\bibitem{Wilde13} M.~Wilde (ALICE Collaboration), Nucl. Phys. A {\bf 904-905}, 573c (2013).

\bibitem{Kaku} M.~Kaku, {\it Quantum field theory: A modern introduction} (Oxford Univ. Press, Oxford, New York, 1993).

\end{thebibliography}
\end{document}